\documentclass[12pt]{article}

\usepackage{xr-hyper}
\usepackage{amsmath, amsfonts, amssymb, amsthm, fullpage, color, bbm, enumerate, titlesec, graphicx, epstopdf, multirow, array, mathtools}
\usepackage{subcaption}
\usepackage[T1]{fontenc}
\usepackage{lmodern}
\usepackage[title]{appendix}

\usepackage[onehalfspacing]{setspace}

\definecolor{darkred}{rgb}{0.5,0,0}

\titleformat*{\paragraph}{\sc}

\usepackage{hyperref}
\hypersetup{allbordercolors={1 1 1},allcolors=darkred,colorlinks=true}
\usepackage[capitalize,nameinlink,noabbrev]{cleveref}

\usepackage[font={small}]{caption}

\usepackage[round]{natbib}
\defcitealias{Chernozhukov2018}{Chernozhukov et al.\ (2018)}
\defcitealias{Chernozhukov2022}{Chernozhukov et al.\ (2022)}

\usepackage[runin]{abstract}

\newcolumntype{C}[1]{>{\centering\arraybackslash}p{#1}}

\theoremstyle{plain}

\crefname{asn}{Assumption}{Assumptions}

\crefname{lem}{Lemma}{Lemmas}

\crefname{prop}{Proposition}{Propositions}

\crefname{cor}{Corollary}{Corollaries}

\newtheorem{lesson}{Lesson}
\crefname{lesson}{Lesson}{Lessons}

\theoremstyle{definition}

\numberwithin{equation}{section}
\numberwithin{figure}{section}
\numberwithin{table}{section}
\numberwithin{lem}{section}
\numberwithin{prop}{section}
\numberwithin{cor}{section}
\numberwithin{asn}{section}

\DeclareMathOperator*{\var}{Var}
\DeclareMathOperator*{\cov}{Cov}

\DeclareMathOperator{\bias}{Bias}
\DeclareMathOperator{\MSE}{MSE}

\DeclareMathOperator{\proj}{proj}

\allowdisplaybreaks

\crefname{sappsec}{Supplemental Appendix}{Supplemental Appendices}
\crefname{sappsubsec}{Supplemental Appendix}{Supplemental Appendices}
\crefname{sappsubsubsec}{Supplemental Appendix}{Supplemental Appendices}
\crefname{appsec}{Appendix}{Appendices}
\begin{document}

\title{\texorpdfstring{\vspace{-1.5\baselineskip}}{} Local Projections or VARs? \texorpdfstring{\\}{}A Primer for Macroeconomists\thanks{Email: {\tt montiel.olea@gmail.com}, {\tt mikkelpm@princeton.edu}, {\tt ericqian@princeton.edu}, and {\tt ckwolf@mit.edu}. Prepared for the NBER Macroeconomics Annual 2025. We are grateful for comments from Isaiah Andrews, Ricardo Caballero, Efrem Castelnuovo, Jim Hamilton, Jonathon Hazell, Ed Herbst, Ethan Ilzetzki, Valerie Ramey, Frank Schorfheide, Jim Stock, Mark Watson, and Iv\'{a}n Werning, and in particular our discussants, Christiane Baumeister and {\`O}scar Jord{\`a}. Plagborg-M{\o}ller acknowledges that this material is based upon work supported by the NSF under Grant {\#}2238049 and by the Alfred P.\ Sloan Foundation, and Wolf does the same for NSF Grant {\#}2314736.}}
\author{\begin{tabular}{ccc}
Jos\'{e} Luis Montiel Olea && Mikkel Plagborg-M{\o}ller \\
{\small Cornell University} && {\small Princeton University} \\[1ex]
Eric Qian && Christian K. Wolf \\
{\small Princeton University} && {\small MIT \& NBER}
\end{tabular}}
\date{\texorpdfstring{\bigskip}{ }May 22, 2025}
\maketitle

\vspace{-1.5em}
\begin{abstract}
What should applied macroeconomists know about local projection (LP) and vector autoregression (VAR) impulse response estimators? The two methods share the same estimand, but in finite samples lie on opposite ends of a bias-variance trade-off. While the low bias of LPs comes at a quite steep variance cost, this cost must be paid to achieve robust uncertainty assessments. Hence, when the goal is to convey what can be learned about dynamic causal effects from the data, VARs should only be used with long lag lengths, ensuring equivalence with LP. For LP estimation, we provide guidance on selection of lag length and controls, bias correction, and confidence interval construction.
\end{abstract}
\emph{Keywords:} dynamic causal effect, impulse response, local projection, misspecification, vector autoregression. \emph{JEL codes:} C22, C32.

\section{Introduction}
\label{sec:intro}

Applied macroeconomists routinely seek to estimate the dynamic causal effects---or impulse response functions---of aggregate shocks to policies or fundamentals. The two dominant empirical methods for doing so are structural vector autoregressions (henceforth interchangeably referred to as VARs or SVARs), going back to \citet{Sims1980}, and local projections (LPs), as introduced by \citet{Jorda2005}. The primary objectives of this paper are, first, to review how to think conceptually about the choice between these two estimation methods, and second, to offer concrete recommendations for applied practice.

We begin with the observation that the choice between LPs and VARs has nothing to do with questions of identification: for any given LP and the economic identifying assumptions that it implements, there exists an equivalent VAR, and \emph{vice versa}. At their core, typical identification schemes in empirical macroeconomics propose to recover the causal effects of aggregate shocks or policies as---sometimes very simple, and sometimes rather complicated---functions of the autocovariances of time series data. Conceptually, LPs and VARs are simply two ways of estimating these autocovariances: they share a common large-sample estimand, and only differ in how they exploit a given finite data set.

In the short samples typical of applied work, the choice between LPs and VARs is one of navigating a bias-variance trade-off. At one end of this spectrum, LPs (or equivalently, VARs with many lags) robustly have low bias, though at the cost of materially elevated variance. At the other end, VARs with few lags tend to deliver sizable precision gains, but at the cost of potentially substantial biases. The intuition is that short-lag VARs extrapolate based only on the first few autocovariances of the data, while LPs flexibly estimate autocovariances at all horizons, without extrapolation. In practice, the variance cost of LPs is so substantial that VARs are typically preferable in terms of mean squared error. The VAR's bias, however, severely threatens the accuracy of its uncertainty assessments, while LP confidence intervals instead accurately reflect statistical uncertainty, by virtue of the LP's small bias. There is ``no free lunch'': if VARs afford \emph{any} precision gains relative to LPs, then their uncertainty assessments are \emph{necessarily} fragile. In fact, because plain LP is semiparametrically efficient, there is actually \emph{no} impulse response estimator that is more efficient than LP in large samples without sacrificing robustness.

Our overall recommendation is that researchers who wish to understand what can be learned about dynamic causal effects from the data---and therefore require confidence intervals with accurate coverage probability---should report results based on LPs, or equivalently based on VARs with many more lags than is currently used in applied practice. Confidence intervals produced by short-lag VARs (or shrinkage techniques such as Bayesian VARs and penalized local projections) are simply too fragile to be trusted. However, if the researcher's goal is merely to forecast or to produce a point estimate of a causal effect for policy analysis, then short-lag VARs or shrinkage techniques are attractive due to their favorable mean squared error in many settings \citep[as emphasized by][]{Li2024}.

We conclude with a ``how-to'' list of practical recommendations for LP estimation and inference. We cover the importance of lag augmentation, selection of control variables and lag length, bias correction, and confidence interval construction.

Due to our focus on simple, concrete take-aways for applied practice, our review of the literature leaves out several specialized or new topics. These include panel data \citep{Almuzara2024}, nonlinear specifications \citep{Caravello2024,Gonccalves2024}, simultaneous confidence bands \citep{MontielOlea2019}, variance decompositions \citep{Plagborg2022}, and certain more technical structural shock identification schemes \citep{Uhlig2005,Baumeister2015}. For brevity, we touch only briefly on proxy or instrumental variable identification, and we abstract from weak instrument issues, even though these are likely to be important in practice \citep{Montiel2021}. Excellent reviews of LPs, VARs, and the relationship between them include \citet{Kilian2017}, \citet{Stock2018}, \citet{Baumeister2024}, and \citet{Jorda2025}.

\paragraph{Outline.}
The organization of the paper follows the outline above: we first review questions of identification, then discuss theoretically and quantify empirically the bias-variance trade-off between LPs and VARs, next ask how to navigate that trade-off in practice, and finally close with a list of concrete practical recommendations. Throughout, our analysis will be deliberately simple, emphasizing clarity over comprehensiveness; for the interested reader we give references to the original literature. To structure our discussion, we summarize our main insights as ``lessons'' that should be viewed as guidelines rather than as formal propositions. Supplementary details---in particular for our empirically calibrated simulations---are provided in the online appendix, and replication code is available online.\footnote{The replication code is available at \url{https://github.com/ckwolf92/lp_var_nberma}.}

\section{Identification}
\label{sec:identification}

In this section we review the basic definition of the LP and VAR estimators, and we establish that they share the exact same estimand (i.e., large-sample limit) when the estimation lag length is large. This equivalence result is nonparametric, and in particular does not require the data generating process to be linear or finite-dimensional. The purpose of economic identifying assumptions is then to establish that this common estimand is actually interesting. The message of this section is that the choice between LPs and VARs is completely orthogonal to structural questions of identification.

\subsection{What do LPs estimate?}
\label{subsec:lps}

We begin with a discussion of the estimand of LPs. LP practitioners seek to estimate the dynamic causal effects of macroeconomic ``shocks'' on aggregate outcomes. To this end, they run linear regressions of the following form, estimated separately by ordinary least squares (OLS) for each horizon $h = 0, 1, 2, \dots$:
\begin{equation}
y_{t+h} = \mu_h + \theta_h^\text{LP} x_t + \gamma_h' r_t + \sum_{\ell=1}^p \delta_{h,\ell}' w_{t-\ell} + \xi_{h,t}. \label{eq:lp_pop}
\end{equation}
They then report as the impulse responses of interest the regression coefficients $\{ \theta_h^\text{LP} \}_h$. In the above regression, $y_t$ is a (scalar) outcome of interest, $x_t$ is a (scalar) impulse variable, $r_t$ is a vector of time series that is included as contemporaneous controls, $w_t = (r_t', x_t, y_t, q_t')'$ collects all variables included as lagged controls, with $q_t$ a potential additional control vector, and $\xi_{h,t}$ is the multi-step forecast error in the regression. Here and in the rest of this section, we consider the limit where the empirical sample size is infinitely large, to abstract from finite-sample issues that will be the focus of the following sections.

By the Frisch-Waugh-Lovell theorem, the regression coefficient $\theta_h^\text{LP}$ has a simple interpretation: it equals the coefficient from a projection of the outcome $y_{t+h}$ on a ``shock'' $\tilde{x}_t$ that is given by the residual from a projection of $x_t$ on the control variables in the regression \eqref{eq:lp_pop}:
\begin{equation} \label{eq:xtilde}
\tilde{x}_t \equiv x_t - \proj ( x_t \mid r_t, w_{t-1},\dots,w_{t-p} ).
\end{equation}

\begin{lesson}
In an LP, we are estimating impulse responses with respect to a shock that is defined as the residual from projecting the impulse variable on the control variables.
\end{lesson}

This LP estimand is economically interesting because, under some assumptions, it can be given a structural interpretation. The argument is most transparent when a researcher directly observes an unpredictable shock (or a valid proxy/instrument for that shock), e.g., by observing the high-frequency responses of an asset price in narrow time windows around policy announcements \citep[as in][]{Kaenzig2021}. In this case, she can straightforwardly use the LP \eqref{eq:lp_pop} to estimate the causal effects of that shock: the impulse variable $x_t$ is the observed shock, the contemporaneous control vector $r_t$ is empty, and finally further controls $q_t$ are not necessary for consistent estimation, but should be included for efficiency and robustness reasons, as discussed in \cref{sec:biasvar_practice}.\footnote{If the researcher is willing to assume an underlying \emph{linear} Structural Vector Moving Average (SVMA) model, then the LP regression coefficients equal the shock's true impulse responses (up to scale). If instead she assumes a more general non-linear causal model, then the estimand is a particular weighted average of marginal effects \citep[see][]{Plagborg2021,Kolesar2024}.} Note that in this case the residualized shock $\tilde{x}_t$ simply equals the shock $x_t$ itself, by virtue of being unpredictable.

In another common class of applications, the impulse variable $x_t$ equals a policy instrument (such as the federal funds rate), the contemporaneous controls $r_t$ equal endogenous variables in the policymaker's reaction function (such as output and inflation), and finally some further lagged controls $q_t$ may be included as well. The residualized shock $\tilde{x}_t$ then equals the disturbance in the policy rule (such as a monetary policy shock) under the classic timing assumption that the endogenous variables $r_t$ do not respond within the period to this disturbance \citep[e.g.,][]{Christiano1999}.\footnote{By defining the shock as the residual in a policy rule where all variables are observed by the econometrician, we are implicitly imposing the assumption of invertibility, i.e., that the policy shock is spanned by current and lagged observed variables. We will comment further on this structural assumption in \cref{subsec:identification_general}.} Intuitively, conditional on the included lagged and contemporaneous controls, the leftover movements in the policy instrument $x_t$ purely reflect shocks to policy, orthogonal to all other disturbances---i.e., exogeneity conditional on controls. We will elaborate on other, more involved ways of identifying macroeconomic shocks in \cref{subsec:identification_general}.

\subsection{What do VARs estimate?}
\label{subsec:vars}

We now consider the second estimation method: VARs. The first step in VAR analysis is to estimate a reduced-form VAR in the vector of all observed time series $w_t$ (using the same definition as above):
\begin{equation}
w_t = c + \sum_{\ell=1}^p A_\ell w_{t-\ell} + u_t. \label{eq:var_pop}
\end{equation}
VAR practitioners translate this reduced-form model into structural impulse response functions using identification conditions. Identifying assumptions are used to map the reduced-form residuals $u_t$ into interpretable structural shocks, and those shocks are then propagated forward through the VAR \eqref{eq:var_pop}. We begin with one popular way of doing so: orthogonalizing the reduced-form residuals $u_t$ using what is known as a recursive ordering. Let $BB' = \Sigma$ be the Cholesky decomposition of the covariance matrix $\Sigma \equiv \var(u_t)$ of the forecast errors $u_t$, where $B$ is a lower-triangular matrix with positive diagonal entries. As in \cref{subsec:lps}, we focus on the response of $y_t$ to the orthogonalized shock to $x_t$. Letting $e_x$ and $e_y$ denote the two unit vectors such that $x_t=e_x'w_t$ and $y_t=e_y'w_t$, the structural VAR impulse response estimate equals
\begin{equation*}
\theta_h^\text{VAR} \equiv e_y'C_h B e_x,
\end{equation*}
where the reduced-form impulse responses satisfy the recursion $C_h = \sum_{\ell=1}^{\min \lbrace h,p \rbrace} A_\ell C_{h-\ell}$, with $C_0$ equal to the identity matrix. Evidently, the VAR impulse responses are obtained by extrapolation: the parameters of the reduced-form VAR \eqref{eq:var_pop} are estimated from observed autocovariances out to lag $p$, and then the parametric structure of the model is exploited to compute responses at all horizons $h$, including $h>p$. 

By standard properties of Cholesky decompositions, this procedure isolates as the ``shock'' the (rescaled) residual in a projection of $u_{x,t} \equiv e_x'u_t$ on $u_{r,t} \equiv e_r'u_t$, where $e_r$ is the unit vector such that $r_t=e_r'w_t$. By definition of the residuals $u_t$, this shock is the same as the residual in a projection of $x_t$ on $r_t$ and $p$ lags of the observables $w_t$---which in turn is precisely the shock $\tilde{x}_t$ in the LP \eqref{eq:lp_pop}.\footnote{In the case of observed shock or proxy variable identification, the VAR specification that we discuss here includes the shock $x_t$ as an ``internal instrument''. The alternative ``external instrument'' VAR procedure \citep[e.g.,][]{Stock2008,Mertens2013} is conceptually distinct and requires the additional assumption of invertibility \citep{Stock2018,Plagborg2021}.} Differently from the LP, however, the impulse response coefficients $\{ \theta_h^\text{VAR} \}_h$ are now not direct projection coefficients of future outcomes $\lbrace y_{t+h} \rbrace_h$ on this common ``shock;'' instead, the dynamic effects of the shock are iteratively propagated forward through the estimated VAR model \eqref{eq:var_pop}.

The simple linear regression logic that we just used to interpret the LP and VAR estimands reveals that the methods are intimately connected: they deliver the dynamic causal effects of the exact same implicit ``shock'' $\tilde{x}_t$, just in one case through direct projection (LP), in the other through iterative propagation (VAR). On impact ($h = 0$), they are thus necessarily the same. Furthermore, as the lag length $p$ becomes arbitrarily large, the extrapolative reduced-form VAR model \eqref{eq:var_pop} is sufficiently flexible to perfectly capture all autocovariance properties of the data, making iterative forecasts equivalent to direct projections; thus, the LP and VAR estimands are identical at \emph{all} horizons $h$ \citep{Plagborg2021,Xu2023}.\footnote{Here we abstract from an inessential issue: the implied LP and VAR shocks may have different variances. But once rescaled to have the same units (e.g., standard deviation equal to 1), equivalence follows.} Though this result pertains to linear estimation methods, we stress that the equivalence is extremely general because it is \emph{nonparametric}, in the sense that it does not impose any particular parametric restrictions (linear or otherwise) on the underlying data generating process. Finally, for general but finite estimation lag length $p$, the extrapolative VAR model still captures the relevant autocovariance properties well out to lag $p$, so that the LP and VAR estimands are typically very close at horizons $h \leq p$, but not necessarily for $h>p$.\footnote{See \citet{Plagborg2021} for further discussion of the finite-$p$ case. As shown there, the LP and VAR estimands at $h \leq p$ for finite $p$ in general differ slightly because the computed impulse responses under LP depend on autocovariances up to lag $p + h$, not just up to $p$. \citet{Ludwig2024} shows that the LP impulse response can be written exactly as a function of VAR impulse responses with varying lag lengths.}

\begin{lesson}
An impulse response from an LP can equivalently be viewed as having been obtained from a VAR that controls for a large number $p$ of lags. In particular, impulse responses from LPs and VARs tend to be very similar at horizons $h \leq p$.
\end{lesson}

\begin{figure}[tp]
\centering
\includegraphics[width=\linewidth]{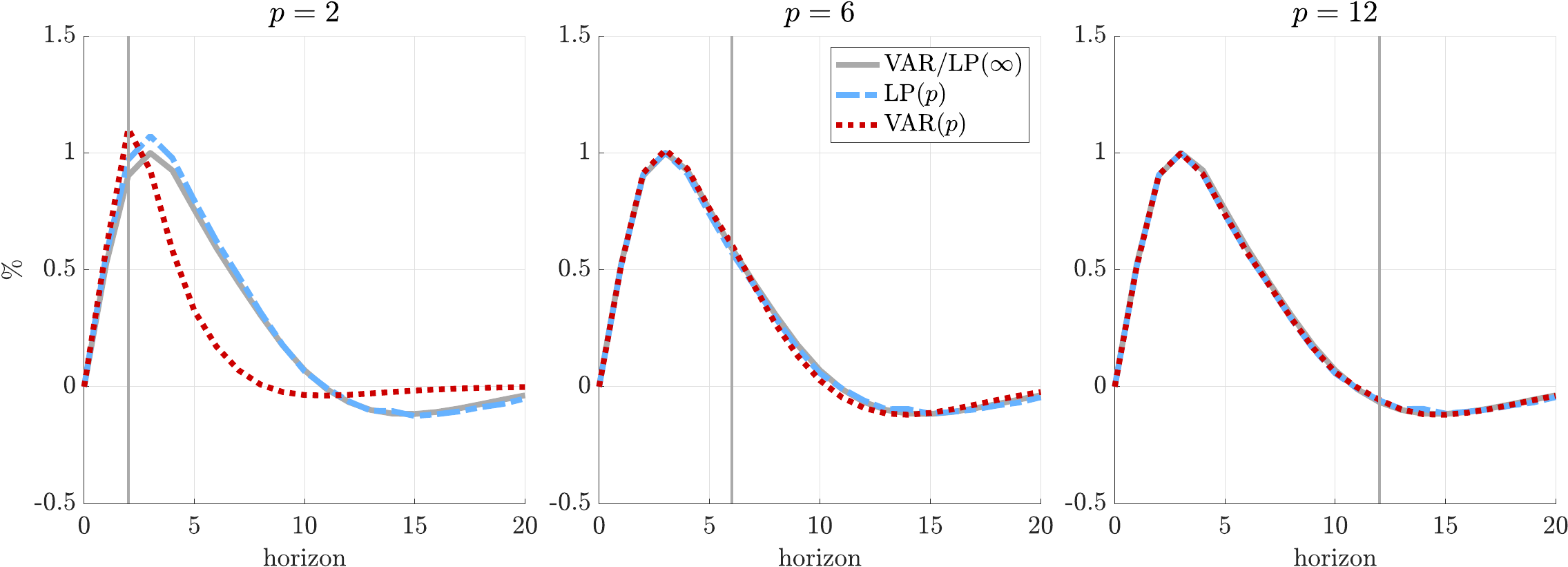}
\caption{LP (blue, dashed-dotted) and VAR (red, dashed) impulse response estimands in the dynamic factor model of \cref{sec:biasvar_practice}. The three panels show the response of output to a recursively identified monetary policy shock, with lag lengths $p = \{ 2, 6, 12 \}$. The grey line is the $p = \infty$ population estimand (for both LP and VAR), and the horizontal line marks the lag length $p$. For further details on the experiment see \cref{app:simulations_estimands}.}
\label{fig:lp_vs_var_dfm}
\end{figure}

We provide a visual illustration of the second lesson in \cref{fig:lp_vs_var_dfm}, which plots LP and VAR impulse response estimands for a recursively identified monetary policy shock, following the classical identification approach of \citet{Christiano1999}. The underlying data generating process (DGP) is the dynamic factor model that we study later in \cref{sec:biasvar_practice}; importantly, that DGP does \emph{not} satisfy a finite-order VAR model. The three panels then plot the VAR and LP estimands for estimation lag lengths $p = \{ 2, 6, 12 \}$ (in red dashed and blue dashed-dotted, respectively), as well as the (common) population estimand ($p = \infty$, in grey). Consistent with our discussion above, LP and VAR estimands are close to each other at horizons $h \leq p$, before then deviating; in particular, for $p$ sufficiently large, they are identical throughout.

\subsection{More general identification approaches}
\label{subsec:identification_general}

The large-sample equivalence between LPs and VARs extends beyond the observed shock and recursive identification schemes discussed above. A common approach to identifying macroeconomic shocks begins with the assumption of ``invertibility'', i.e., that the true shocks are spanned by current and lagged time series observables $w_t$ \citep[see][for a standard treatment]{Fernandez2007}. Identification schemes like the recursive ordering of \citet{Christiano1999}---which, as we showed above, can equivalently be implemented using either LPs or VARs---combine the assumption of invertibility with short-run timing restrictions on the shocks. Invertibility may however also be combined with other restrictions to identify shocks, including for example long-run restrictions \citep{Blanchard1989} or sign restrictions \citep{Uhlig2005}. In all these identification approaches, the structural shock of interest ultimately equals $\beta'u_t$, where $u_t$ is the reduced-form forecast error from \eqref{eq:var_pop}, and the weight vector $\beta$ is a (potentially complicated) function of the researcher's identifying assumptions. A structural VAR practitioner would then report the impulse response estimate $\theta_h^\text{VAR}=e_y'C_h \Sigma \beta$. But by our earlier Frisch-Waugh-Lovell logic, we can alternatively compute impulse responses with respect to the same shock $\beta'u_t$ from the more general LP
\begin{equation}
y_{t+h} = \mu_h + \theta_h^\text{LP} (\beta' w_t) + \sum_{\ell=1}^p \delta_{h,\ell}' w_{t-\ell} + \xi_{h,t}. \label{eq:lp_pop_richid}
\end{equation}
Intuitively, even such richer approaches to identification ultimately amount to the researcher giving a causal interpretation to some (perhaps complicated) function of the autocovariance function of the observed data. Either VARs or LPs can then be used to compute this shared estimand of interest, with the two methods necessarily agreeing in large samples when the estimation lag length $p$ is large.\footnote{See \citet{Plagborg2021} for a detailed discussion of how exactly identification schemes like long-run and sign restrictions map into $\beta$.}

\begin{lesson}
When it comes to identification, anything you can do with VARs, you can do with LPs, and \emph{vice versa}.
\end{lesson}

In conclusion, the choice between LPs and VARs has absolutely nothing to do with the identifying assumptions necessary to isolate a given shock of interest. Instead, the question is simply which method is better at (i) recovering this common estimand in finite samples and (ii) quantifying the associated statistical uncertainty. This is why \emph{estimation} and \emph{inference}---and not \emph{identification}---will be the focus of the remainder of the paper.

\section{The bias-variance trade-off}
\label{sec:bias_variance}

We now dig deeper into the econometric properties of LPs and VARs. First, we motivate our analysis by demonstrating that, despite the population equivalence between LPs and VARs when the estimation lag length is large, the choice between the two estimators does matter in practice when using a small or moderate lag length, as is typical in applied work. Second, we present simple illustrative simulations of the properties of LP and VAR estimators. Finally, we review the available econometric theory. The overarching takeaway of the analysis will be that, in finite data sets, there is a clear and inevitable bias-variance trade-off between LPs and (short-lag) VARs: small bias and large variance for LPs, and \emph{vice versa} for VARs. In fact, because plain LP is semiparametrically efficient, \emph{no} estimator can outperform it in terms of variance without sacrificing robustness.

\subsection{VARs vs.\ LPs in empirical work}

We first establish that, in finite samples and with a moderate estimation lag length, LPs and VARs can provide meaningfully different estimates of their common estimand (that would, as we saw, obtain in large samples if we were able to control for very many lags). Our analysis is based on the literature synthesis of \citet{Ramey2016}. We replicate four of the headline applications of that paper, for shocks to monetary policy, taxes, government purchases, and technology, respectively. We then use LPs and VARs to estimate impulse responses for several variables and at several horizons, throughout staying as close as possible to the specifications considered by \citeauthor{Ramey2016}. Across all response variables and horizons, we then compare standard errors and compute the differences in point estimates. Implementation details are provided in \cref{app:varlp_emp}.

\begin{figure}[tp]
\centering
\includegraphics[width=\linewidth]{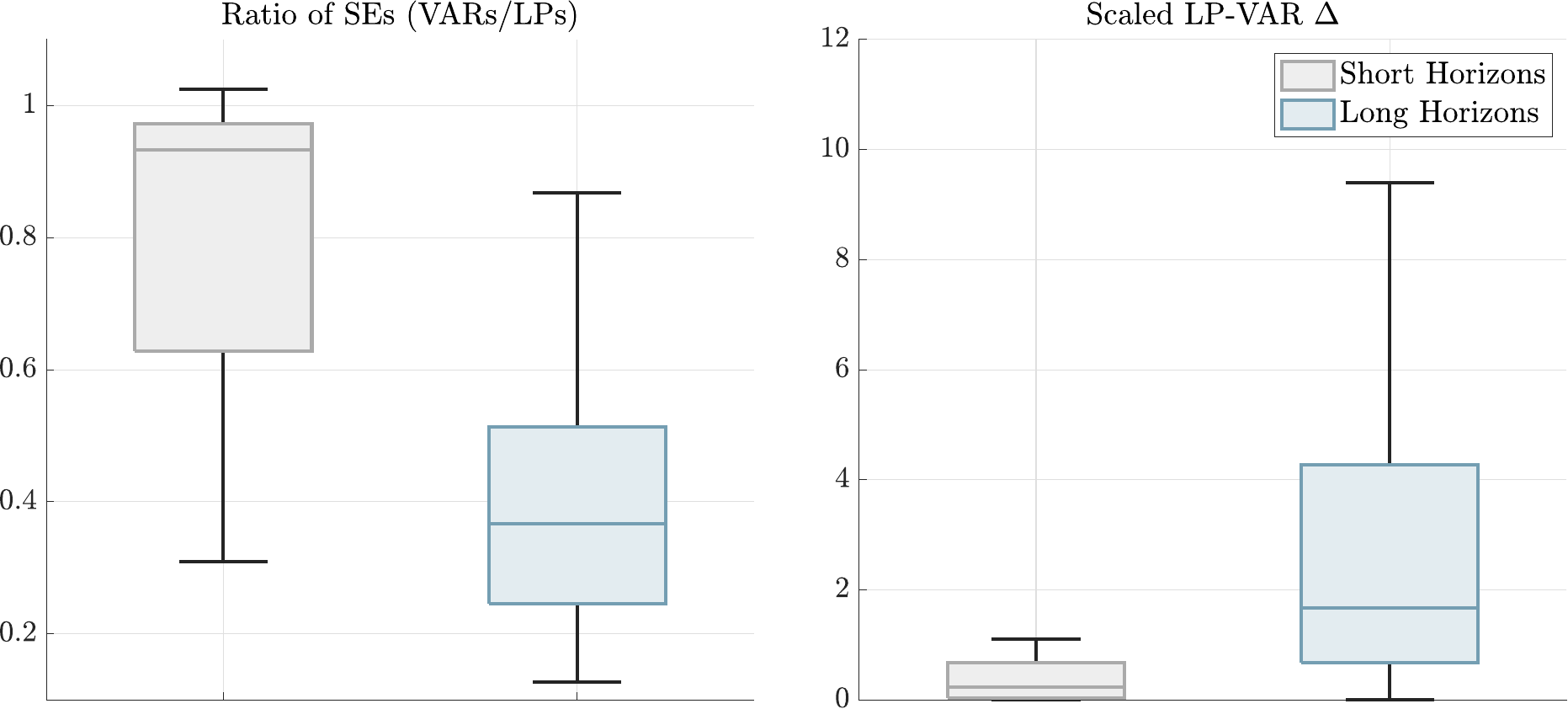}
\caption{Box plots of VAR-to-LP standard error ratios (left panel) and normalized point estimate differences (right panel) for short horizons (blue) and for long horizons (grey). The central mark indicates the median, the bottom and top edges of the box indicate the 25th and 75th percentiles, and the whiskers extend to the most extreme data points that are not more than 1.5 times the interquartile range away from the bottom or top of the box (i.e., ignoring outliers). Based on applications in \citet{Ramey2016}. See \cref{app:varlp_emp} for details.}
\label{fig:lp_vs_var_ramey}
\end{figure}

\cref{fig:lp_vs_var_ramey} shows that there are meaningful differences in both precision as well as location of the two estimators: VAR impulse responses often have substantially lower standard errors than LPs (left panel), and the two sets of point estimates can be quite far apart (right panel). In the left panel we display a box plot of the ratio of VAR and LP standard errors across all different shocks and outcome variables, separately for short horizons (blue, $\leq$ one year) as well as long horizons (grey, $>$ one year). At short horizons (i.e., for $h$ close to or below the lag length), VARs are only somewhat more precise than LPs, with the median standard error ratio only slightly below one, consistent with our theoretical discussion in \cref{sec:identification}.\footnote{Asymptotically, VAR standard errors are weakly smaller than LP standard errors, so the standard error ratio is bounded above by $1$, as we will explain later. This need not be the case in small samples, however.} At long horizons, VARs are instead materially more precise, yet again as expected given the preceding discussion. The right panel then shows the difference in point estimates between the two methods, normalized by the VAR standard error. We see that differences can often be of much greater magnitude than the VAR standard error. The gaps are instead smaller at short impulse response horizons, as expected.

\begin{lesson}
At intermediate and long horizons, LP and VAR impulse response estimates in empirical applications are often materially different, and VAR estimates typically have much lower standard errors than LP. 
\end{lesson}

These empirical results can of course not directly tell us whether the LP or the VAR estimates tend to be closer to the truth. For this, we will next turn to a simulation exercise in a simple data generating process, before presenting the general econometric theory.

\subsection{A numerical illustration}

As a first step toward gaining intuition for the econometric properties of LPs and VARs, we present some illustrative simulations. For pedagogical reasons, we here assume a simple univariate ARMA$(1,1)$ DGP, while leaving empirically realistic simulations to later sections:
\begin{equation} \label{eq:sim_arma}
y_t = \rho y_{t-1} + \varepsilon_t + \alpha \varepsilon_{t-1},\quad \varepsilon_t \overset{i.i.d.}{\sim} N(0,\sigma^2),
\end{equation}
where $y_t$ is an observed scalar outcome variable, $\varepsilon_t$ is an unobserved scalar shock, and $\rho$ and $\alpha$ are parameters. We are interested in the impulse response of $y_t$ with respect to $\varepsilon_t$ at horizon $h$, which has the formula $\theta_h \equiv \rho^h + \alpha \rho^{h-1}$ for $h \geq 1$ (and $\theta_0=1$).

\begin{figure}[tp!]
\centering
\includegraphics[width=.75\textwidth]{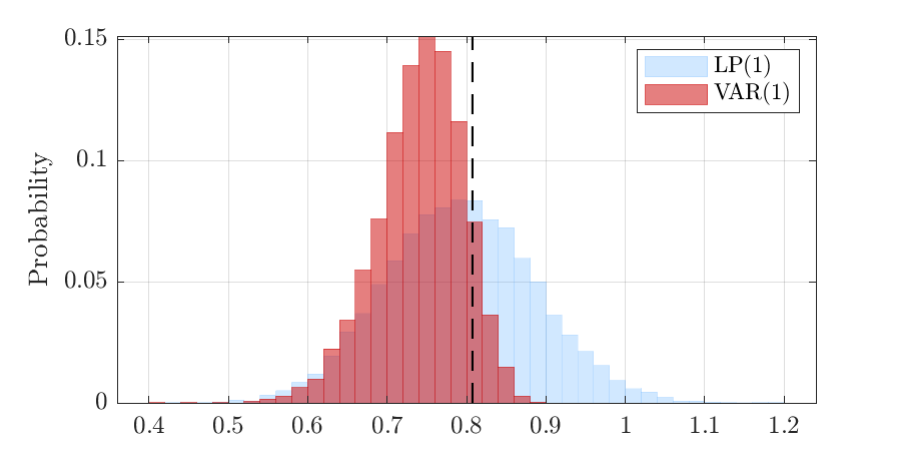}
\caption{Histogram of VAR(1) (in red) and LP(1) (in blue) impulse response estimates for $h=2$, $\rho=0.85$, and $\alpha=0.1$ relative to the true impulse response $\theta_h$ (dashed line).}
\label{fig:arma_lpvar_dens}
\end{figure}

\cref{fig:arma_lpvar_dens} shows that, in this DGP, there is a stark bias-variance trade-off between LPs and (short-lag) VARs. The figure shows histograms of the LP (blue) as well as VAR (red) estimators of the impulse response at horizon $h=2$ for $\rho=0.85$, $\alpha=0.1$, and with a sample size of $T=240$. Both estimators control for one lag of the data, with larger lag lengths to be considered below.\footnote{The VAR estimator uses an AR(1) specification. The LP estimator regresses $y_{t+h}$ on $y_t$, while controlling for $y_{t-1}$. Both estimators include an intercept and omit the bias corrections discussed in \cref{sec:lpvar_details}.} We can see that the LP estimator is centered close to the true value of the impulse response, $\theta_h$. Intuitively, because LP just directly projects the future outcome of interest $y_{t+h}$ on $y_t$ (controlling for $y_{t-1}$), it is able to pick up on the full ARMA dynamics of the DGP \eqref{eq:sim_arma}. The VAR estimator, in contrast, suffers from extrapolation bias: its impulse response estimate $\hat{\rho}^h$ is obtained by first estimating the autoregressive parameter $\hat{\rho}$ through a \emph{one}-period-ahead forecast regression of $y_t$ on $y_{t-1}$, and then iterating forward $h$ steps using the parametric AR(1) model. The VAR estimator therefore is not directly informed by the sample autocovariances at horizons 2 and longer, causing it to miss out on the more intricate moving average dynamics present in the DGP. Yet, the figure also shows that the parametric extrapolation has a clear benefit: because the first-order autoregressive coefficient $\hat{\rho}$ is more precisely estimated than the longer-horizon autocovariances, the sampling distribution of the VAR impulse response estimator is less dispersed than the LP estimator. In summary, we see a trade-off: the sampling distribution of the LP estimator is well-centered but dispersed, while that of the VAR estimator is more tightly concentrated but centered incorrectly.

\begin{figure}[t!]
\centering
\begin{subfigure}[c]{0.49\textwidth}
\includegraphics[width=\textwidth]{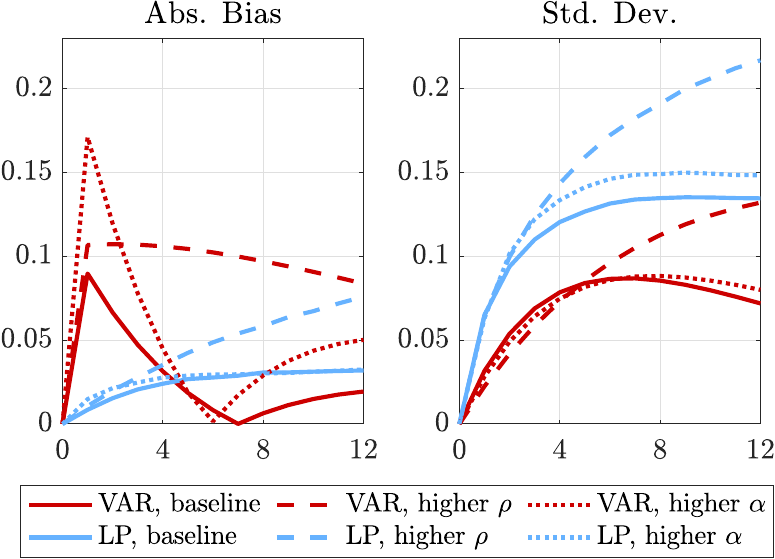}
\caption{\textsc{Alternative DGPs}}
\end{subfigure}
\begin{subfigure}[c]{0.49\textwidth}
\includegraphics[width=\textwidth]{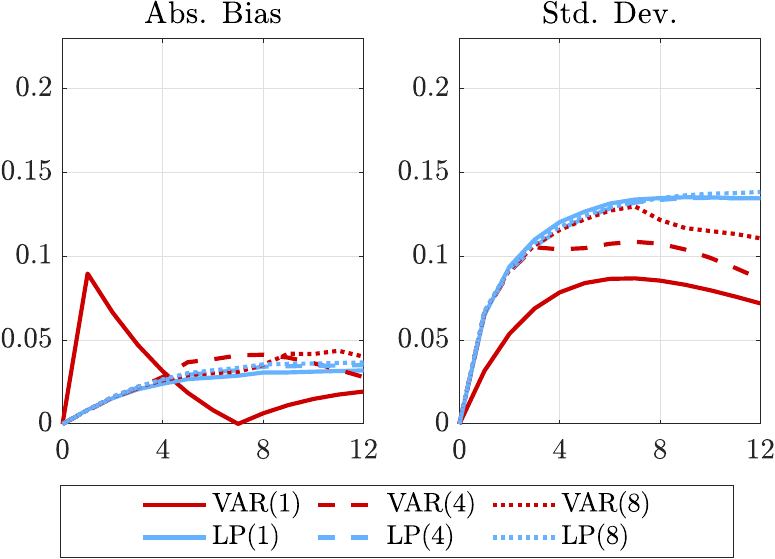}
\caption{\textsc{Alternative lag lengths}}
\end{subfigure}
\caption{Absolute bias and standard deviation of LP and VAR estimators, as a function of horizon $h$. In the left panel, VAR and LP are estimated with 1 lag.  Compared to the baseline DGP ($\rho=0.85$ and $\alpha=0.1$), ``higher $\rho$'' takes $\rho=0.95$  and ``higher $\alpha$''  takes $\alpha=0.2$. The right panel varies the estimation lag length $p$ for the baseline DGP.}
\label{fig:arma_biasvce}
\end{figure}

Panel (a) of \cref{fig:arma_biasvce} exhibits how the nature of the bias-variance trade-off differs across DGPs and impulse response horizons. On impact, $h=0$, the two estimators are numerically identical (they both equal 1). However, at intermediate horizons the bias-variance trade-off is stark, for the reasons discussed earlier. The larger the moving average coefficient $\alpha$, the more misspecified is the one-lag VAR specification, and consequently the larger is the VAR bias. The VAR bias then decreases at long horizons, since in stationary DGPs the impulse responses must converge to zero as $h \to \infty$, a feature that is mechanically enforced by the VAR estimator $\hat{\rho}^h$ (but not by the LP estimator). However, the figure also reveals that the speed of convergence of the impulse responses to zero depends on the persistence parameter $\rho$, so it is not clear \emph{a priori} at what horizons we should expect the impulse responses to be small---and thus when we can stop worrying about the VAR bias.

Finally, panel (b) of \cref{fig:arma_biasvce} shows how changes in the estimation lag length $p$ shape the bias-variance trade-off, demonstrating that the bias of the VAR estimator at shorter horizons is reduced when the lag length is increased. In this panel, the LP and VAR estimators now control for $p$ lags of the data $y_{t-1},\dots,y_{t-p}$ rather than just one. When we control for $p=4$ lags, the VAR impulse response estimator is now nearly unbiased out to horizon $h=4$. This comes at a cost, however: the VAR variance increases substantially, to essentially equal that of LP at those horizons. When we control for even more lags, the VAR bias is now reduced at even longer horizons, but again at the cost of higher variance. These results hark back to our second lesson: an LP is essentially a VAR that controls for a lot of lags. Consistent with this interpretation, the bias and standard deviation of the VAR estimator depend dramatically on the choice of lag length, while this choice matters much less for the LP estimator.

In the next section we will establish that these simulation-based conclusions in fact hold up theoretically in a much wider class of multivariate models. Subsequent sections will then use large-scale, empirically calibrated simulation studies to demonstrate that these takeaways are practically relevant, in addition to being theoretically well-founded.

\subsection{Theoretical insights}
\label{subsec:bias_variance_analytics}

To gain a deeper understanding of the bias-variance trade-off, we review some of the main theoretical results from \citet{Montiel2024}.

\paragraph{Univariate model.}
We begin with the univariate ARMA$(1,1)$ model \eqref{eq:sim_arma}. \citeauthor{Montiel2024} show that the VAR and LP estimators of the true impulse response $\theta_h$ are both approximately normally distributed in large samples, though with differing bias and variance:
\begin{equation} \label{eq:large_sample_distr}
\hat{\theta}_{h}^{\text{VAR}} \overset{\cdot}{\sim} N\left(\theta_h + b_h(p), \tau_{h,\text{VAR}}^2(p)\right),\quad \hat{\theta}_{h}^{\text{LP}} \overset{\cdot}{\sim} N\left(\theta_h, \tau_{h,\text{LP}}^2\right),
\end{equation}
where $\overset{\cdot}{\sim}$ denotes ``approximate distribution in large samples'' (in the usual formal sense of asymptotic normality), $b_h(p)$ is a VAR bias term that depends on the estimation lag length $p$, and the large-sample variances satisfy $\tau_{h,\text{LP}}^2 \geq \tau_{h,\text{VAR}}^2(p)$; we will return later to the observation that the LP large-sample variance does not depend on the estimation lag length.

The formal result that yields \eqref{eq:large_sample_distr} requires the moving average coefficient $\alpha$ in the DGP \eqref{eq:sim_arma} to be relatively small. Thus, our analysis gives the VAR estimator the benefit of the doubt by viewing the true DGP as close to---but not exactly equal to---an AR(1), in a way that yields a non-trivial trade-off between model misspecification and statistical sampling error.\footnote{Formally, $\alpha$ needs to be proportional to the magnitude of the standard deviation of the estimators (i.e., $\alpha=\alpha_T \propto T^{-1/2}$). If we instead analyzed the asymptotic properties of fixed-lag VAR estimators under the DGP with fixed parameter $\alpha$ (as in \citealp{Braun1993}), the conclusions would be stark but empirically uninteresting: for any moving average coefficient $\alpha \neq 0$, the VAR estimator would be inconsistent for the true impulse response due to misspecification, so for large sample sizes the ratio of the bias to the standard deviation of the estimator diverges to infinity, yielding a trivial bias/variance trade-off in the limit. The ``local-to-zero'' modeling device of setting $\alpha \propto T^{-1/2}$ \citep[which follows][]{Schorfheide2005} should not be viewed as a literal description of reality, but rather as a technical device intended to tractably and accurately capture key finite-sample phenomena, similar to the econometric literatures on weak instruments or near-unit roots. Our later simulations will show that the lessons learned from these local-to-zero asymptotics are borne out in a large class of realistic DGPs.} The popularity and empirical success of autoregressive estimators in the forecasting literature suggest that carefully implemented autoregressive specifications are often not grossly misspecified. However, it seems unlikely that such simple parametric models can successfully capture every aspect of the real world. These observations make our setting a natural one to evaluate the properties of VAR and LP estimators.

We now study in greater detail the bias and variance expressions in \eqref{eq:large_sample_distr}.

\begin{itemize}

\item \emph{Bias.} The first important implication of \eqref{eq:large_sample_distr} is that, while the LP estimator has approximately zero bias in large samples, the VAR estimator is generally biased. When a single lag $p=1$ is used for estimation, the VAR bias has the simple formula
\begin{equation*}
b_h(1) = \underbrace{h\rho^{h-1}}_{\frac{\partial (\rho^h)}{\partial \rho}} \underbrace{(1-\rho^2)\alpha}_{\approx \frac{\cov(y_{t-1},\alpha \varepsilon_{t-1})}{\var(y_{t-1})}} - \underbrace{\rho^{h-1}\alpha}_{\theta_h-\rho^h}.
\end{equation*}
The product of the first two factors on the right-hand side is due to the endogeneity bias in the estimated autoregressive coefficient $\hat{\rho}$, here caused by the lagged error term $\alpha\varepsilon_{t-1}$ in the regression of $y_t$ on $y_{t-1}$. The second factor is the bias of $\hat{\rho}$, while the first factor maps $\rho$ into the impulse response of interest. The last term is due to the VAR estimator then relying on a misspecified functional form to transform $\hat{\rho}$ into impulse responses $\hat{\rho}^h$. It can be shown that both these sources of bias can be reduced to zero by sufficiently increasing the estimation lag length $p$; however, as we will see, this necessarily comes at the cost of higher variance.

The key to the zero asymptotic bias of the LP is that we control for a lag $y_{t-1}$ of the data. Intuitively, by controlling for a lag of the data, we are effectively regressing $y_{t+h}$ on the residualized ``shock'' $\tilde{x}_t=y_t - \proj(y_t \mid y_{t-1})$. Since this residualized shock correlates only weakly with further lags of the data under the ARMA model \eqref{eq:sim_arma} with small $\alpha$, the textbook omitted variable bias formula for regression reveals that the LP bias will be negligible. Increasing the lag length even further has no effect on the asymptotic bias, which remains zero; in contrast, an LP estimator that regresses $y_{t+h}$ on $y_t$ (i.e., without a lagged control) would be subject to bias of the same magnitude as its standard deviation (like the VAR), and so should be avoided.

\item \emph{Variance.} While LP dominates VAR in terms of bias, the large-sample variance $\tau_{h,\text{LP}}^2$ of LP necessarily exceeds the variance $\tau_{h,\text{VAR}}^2(p)$ of the VAR, and typically strictly so. Intuitively, since we assume the model to be rather ``close'' to a finite-order VAR (i.e., only mild misspecification), the large-sample variance of LPs and VARs can be shown to be the same as in a correctly specified VAR model. In that correctly specified model, however, the VAR estimator is efficient by virtue of being the quasi-maximum-likelihood estimator. The high variance of LP can easily be spotted in practice: LP-estimated impulse response functions tend to look much more jagged or erratic (as a function of the horizon) than VAR-estimated impulse response functions. This jaggedness is the price to pay for low bias. Increasing the LP lag length has no further effect on the asymptotic variance; increasing the VAR lag length instead increases the variance, with the VAR's asymptotic bias disappearing only when its impulse responses are just as jagged and volatile as those of LP.

\end{itemize}

To summarize, in the simple ARMA$(1,1)$ model \eqref{eq:sim_arma} considered so far, LPs and VARs indeed lie on opposite ends of a bias-variance trade-off, with this trade-off vanishing as the VAR estimation lag length is increased, consistent with our earlier simulations.

\paragraph{Multivariate generalization.}
\citet{Montiel2024} show that all the above-mentioned qualitative lessons go through in a much wider class of multivariate VARMA$(p_0,q_0)$ models, where $p_0$ is some finite true autoregressive order, while the moving average lag length $q_0$ could be infinite. As before, this set-up gives the benefit of the doubt to VAR practitioners by modeling the moving average coefficients as relatively ``small.'' Importantly, this class is general enough to be empirically relevant, for two main reasons. First, the class is consistent with typical linearized structural macroeconomic models.\footnote{Linearized DSGE models in macroeconomics almost always have a VARMA representation, though they typically do not have an exact finite-order VAR representation.} Second, it captures several important kinds of dynamic misspecification: even if the true DGP were a finite-order VAR, if the econometrician accidentally omits some lags or relevant control variables from the empirical specification, then those omissions will show up as moving average terms \citep{Granger1976}.

It turns out that, even in this much larger class of models, the large-sample distributions \eqref{eq:large_sample_distr} continue to apply, and thus so does the bias-variance trade-off, as long as we consider LP and VAR estimators that control for at least $p \geq p_0$ lags of the data; if we fail to control for at least $p_0$ lags, either estimator could be badly biased in large samples. The way to interpret this result in practice is that it is important to control for those variables and lags that are \emph{strongly} predictive of \emph{either} the outcome variable of interest \emph{or} the impulse variable; in the present theoretical setting, this amounts to controlling for $p_0$ lags of the data vector, since the moving average coefficients are small relative to the autoregressive coefficients. But once we control for all of these strong predictors, the LP estimator has zero asymptotic bias and constant variance as a function of $p$, even though further lags of the data do have some modest remaining predictive power. In other words, LP is relatively insensitive to the omission of moderately important control variables and lags (modeled through the moving average terms). By contrast, the VAR estimator has low variance yet generally nonzero bias even once we control for the most important predictors; intuitively, the issue is that the VAR does not project on the most important predictors horizon-by-horizon (like LP), but instead only does so on impact, and then iterates forward using its particular parametric structure.

\begin{lesson}
In finite samples, there is a stark bias-variance trade-off between LPs and VARs with small-to-moderate lag lengths. VARs extrapolate, and so have low variance but potentially high bias. LPs instead do not extrapolate and so have low bias and high variance.
\end{lesson}

There is of course a simple way to ``bias-correct'' the VAR estimator, as already discussed: use a very large estimation lag length. In particular, in the present model set-up, it can be shown that the large-sample bias $b_h(p)$ of the VAR estimator at horizon $h$ is zero when the estimation lag length $p$ exceeds $p_0+h$. However, when the lag length is chosen this large, then the variance is necessarily inflated to that of LP---in other words, the VAR bias is zero \emph{only because} long-lag VAR and LP estimators are equivalent in large samples. In practice, we do not know exactly how large the true autoregressive lag length $p_0$ is, so to be certain that we included enough lags to eliminate meaningful biases, we would need to compare the VAR results with comparable LP results and verify that they are close to each other.

\paragraph{No free lunch for VARs.}
As the final step in our theoretical analysis we show that even relatively minor amounts of VAR misspecification can yield economically large biases. This discussion will rationalize much of what we find in our later simulations.

Our approach here is to ask how large the VAR bias $b_h(p)$ can possibly be, as a function of the magnitude of the dynamic misspecification. \citet{Montiel2024} show that the ratio of the absolute bias to the standard deviation of the VAR estimator is (asymptotically) bounded by
\begin{equation} \label{eq:bias_bound}
\frac{|b_h(p)|}{\tau_{h,\text{VAR}}(p)} \leq \sqrt{T \times \mathcal{M}} \times \sqrt{\frac{\tau_{h,\text{LP}}^2}{\tau_{h,\text{VAR}}^2(p)}-1},
\end{equation}
where $T$ is the sample size and $\mathcal{M}$ is a measure of the overall magnitude of the misspecification of the VAR, namely the fraction of the variance of the moving average residual that is explained by the lagged shocks in the VARMA model. Equation \eqref{eq:bias_bound} states that the VAR's bias is bounded above by a simple function of just two numbers: the potential amount of misspecification, and the ratio of LP-to-VAR standard errors. In a well-specified finite-order VAR model there are no lagged shocks that enter into the residual, and so in this case there is no misspecification, i.e., $\mathcal{M}=0$, and the bias is zero. In the simple ARMA$(1,1)$ model \eqref{eq:sim_arma}, $\mathcal{M} = \alpha^2$. For a concrete example, suppose that $\mathcal{M}=0.01$, so that the lagged shocks account for a mere 1\% of the variance of the residual, and furthermore assume that $T=100$ and the VAR standard error equals 0.4 times the LP standard error, $\tau_{h,\text{VAR}}(p)/\tau_{h,\text{LP}}=0.4$, roughly consistent with the median standard error ratio at longer horizons in \cref{fig:lp_vs_var_ramey}. Then the upper bound \eqref{eq:bias_bound} equals $\sqrt{5.25} \approx 2.29$, so that the bias of the VAR estimator can be more than twice as large as its standard error, despite the minor degree of misspecification in this example. Biases of this magnitude are obviously very worrisome when drawing statistical inferences from conventional VAR regression output. Though the formula \eqref{eq:bias_bound} represents an upper bound on the extent of the bias, it can be shown that there always exists a residual moving average process (with small coefficients satisfying the imposed misspecification magnitude $\mathcal{M}$) that achieves this bound. Moreover, the particular form of this ``least favorable'' residual process does not appear to be unreasonable \emph{ex ante} based on economic theory, and conventional statistical tests have low power to detect it \emph{ex post} in the data.

To summarize, the worst-case bias formula \eqref{eq:bias_bound} implies that the \emph{only} way for VAR practitioners to \emph{guarantee} that the bias is negligible is to include so many lags $p$ in the specification that the standard error becomes equal to that of LP. Conversely, for any VAR specification that delivers material efficiency gains relative to LP, there is reason to worry that the VAR bias could be large relative to the standard error.

\begin{lesson}
There is no free lunch for VAR practitioners: whenever VARs have small standard errors relative to LP, there is good reason to worry about potentially large biases due to model misspecification. Conventional model selection procedures or specification tests do not adequately guard against this risk. The only way to ensure that VAR impulse responses have low bias is to control for so many lags that they become equivalent with LP.
\end{lesson}

The absence of a free lunch has particular bite at long horizons. If the DGP is stationary, impulse responses must be close to zero at very long horizons, but it is rare to possess precise prior knowledge about exactly at what rate the impulse responses decay. Unless the lag length is very large, the VAR estimator simply extrapolates the long-run responses based on the short-run empirical autocovariances, eventually enforcing an exponential rate of decay of the impulse responses in stationary environments. The VAR-estimated impulse responses at long horizons will therefore tend to have very small standard errors relative to LP (which, as we have discussed, does not enforce exponential decay). But this, then, is precisely when the potential for VAR bias will also be high: if the estimated rate of decay is inaccurate, this can dramatically affect the magnitude of estimated long-horizon responses, and their bias can be several times larger than their standard errors, as indicated by the bias bound \eqref{eq:bias_bound}. In turn, this can lead to large errors in estimating such key features as the half-life or quarter-life of the impulse response function, or its cumulative value (as would be relevant for estimating hysteresis effects, say). LPs remain approximately unbiased at long horizons, and their large standard errors accurately reflect the fundamental issue that any finite data set only has limited information about what happens in the long run, in the absence of prior information. In particular, LPs simply do not allow the estimation of impulse responses at ultra-long horizons $h$ beyond the observed sampled size $T$ (or rather, the effective sample size $T-p$); VARs instead produce such estimates by pure extrapolation, with potential for severe bias. The fact that an LP at horizon $h$ uses only $T-p-h$ data points (as opposed to $T-p$ for a VAR) is not a drawback of the procedure; it is a necessary consequence of the desire to avoid the extrapolation that the VAR estimator relies on.

\subsection{Trade-offs between LPs and other estimators}
\label{subsec:biasvar_general}

In addition to VARs, the literature has proposed several variants of and alternatives to local projection estimators intended to increase efficiency, including for example Generalized Least Squares (GLS) variants of LP, Autoregressive Distributed Lag (ADL) regressions, and parametrized LPs. This goal, however, faces one key theoretical obstacle: LPs are semiparametrically efficient and therefore cannot be improved upon without sacrificing robustness or imposing substantive further restrictions. In this sense, the bias-variance trade-off between LPs and VARs extends also to comparisons with other estimation procedures.

The plain OLS LP estimator is semiparametrically efficient in linear models. This was initially conjectured by \citet{Plagborg2021} based on the equivalence of the LP and VAR estimands (recall \cref{sec:identification}), with \citet{Xu2023} offering the formal proof. Specifically, in a general dynamic linear model with unrestricted lag structure, the efficient quasi-maximum likelihood estimator is a VAR estimator with an asymptotically increasing lag length.\footnote{More precisely, a VAR estimator with asymptotically increasing lag length reaches the semiparametric efficiency bound associated with the conditional moment restrictions implied by a VAR($\infty$) model with shocks that are martingale difference sequences. \citet{Xu2023} requires the model to be stationary with homoskedastic shocks, but it seems likely that these assumptions can be relaxed.} The OLS LP estimator (also with an increasing lag length) is asymptotically equivalent with this estimator. Thus, in this setting, any well-behaved impulse response estimator must \emph{either} have weakly higher variance than plain LP in large samples, \emph{or} it must be inconsistent in some DGPs where plain LP is consistent. In other words, any efficiency gains can only come at the expense of imposing substantive restrictions on the transmission mechanism or the identification of the shock. Of course, such additional restrictions may very well be plausible in some applications, but they should always be acknowledged.

\begin{lesson}
If an estimator achieves lower variance in large samples than plain LP, it must require extra restrictions on economic transmission mechanisms and/or shock identification.
\end{lesson}

This discussion has immediate implications for several of the recently proposed alternatives to the plain OLS LP estimator. First, GLS variants of LP cannot uniformly improve on \citeauthor{Jorda2005}'s (\citeyear{Jorda2005}) original OLS version. In particular, \citet{Xu2023} shows that the variance gains of the GLS approach of \citet{Lusompa2023} only obtain when the true DGP is a VAR with short lag length.\footnote{In fact, some implementations of GLS LP \citep[e.g.,][]{Breitung2023} are equivalent in large samples with a VAR estimator and therefore subject to the same bias issues discussed earlier, unless the estimation lag length is large.} Second, ADL regressions featuring the shock $x_t$---effectively an LP that controls additionally for \emph{future} values $x_{t+1},\dots,x_{t+h}$ of the shock $x_t$---can provide efficiency gains, as they exploit the information that the shock is directly observed \citep{Choi2019,Baek2022}. Intuitively, because the shock is serially uncorrelated, future shocks are valid controls and help increase efficiency, as discussed further in \cref{sec:lpvar_details}. However, a crucial caveat is that, if the serially uncorrelated shock is not in fact directly observed but needs to be pre-estimated, then the ADL standard errors must be adjusted to reflect the generated regressors. Not only is this complicated to do in practice, it may also negate any efficiency gains relative to plain LP. Indeed, no ADL procedure with pre-estimated shocks can outperform plain LP for general shock identification procedures. Third, parametrized LPs impose shape restrictions on the impulse response functions \citep[e.g.,][]{Barnichon2018}. Such restrictions will increase precision but in general also lead to bias.

\subsection{Outlook}

In \cref{sec:identification,sec:bias_variance} we have made some broad conceptual points on identification as well as on the bias-variance trade-off between LPs, VARs, and related techniques. In the remainder of the paper we give recommendations on how to navigate this trade-off in empirical practice. \cref{sec:lpvar_details} begins with specification choices and implementation details for a range of LP and VAR estimators. \cref{sec:biasvar_practice} then provides a quantitative, empirically calibrated assessment of the bias-variance properties of these estimators, while \cref{sec:uncertainty} discusses implications for inference. Finally \cref{sec:recommend} concludes with our recommendations for applied practice.

\section{LP and VAR specification choices}
\label{sec:lpvar_details}

Here we discuss three of the main practical considerations involved in estimating impulse responses through either LPs or VARs: selection of control variables and lags, bias correction, and whether to shrink the estimates towards \emph{a priori} plausible values.

\subsection{Control variables and lag length}
\label{subsec:lags_controls}

In applications of LP or VAR estimators, a ubiquitous question is which variables and how many lags we should control for. There are four main factors to consider.

\begin{enumerate}

\item The controls should ensure valid identification of an interpretable shock. As discussed in \cref{sec:identification}, both LPs and VARs ultimately project the outcomes of interest on a measure of a shock, which by the Frisch-Waugh-Lovell theorem is simply given by the residual after regressing the impulse variable on the full list of controls. That list of controls, and the number of lags of these, should be sufficiently rich such that the residualized shock is unpredictable from other external variables or further lags of the data. In applications of recursive/Cholesky identification, the residualized shock often can be directly interpreted as the residual in a policy rule (e.g., a Taylor rule residual), and so economic theory can be brought to bear on what to control for (e.g., variables in the central bank's reaction function). If the impulse variable is a credibly unpredictable ``shock,'' for example because it is obtained from high-frequency asset price movements over short time windows around policy announcements, then control variables are not needed to ensure correct identification. However, even in this case it is important to include controls for the reasons outlined next.

\item The second purpose of control variables is to increase efficiency. As in a randomized control trial, even when control variables are not correlated with the treatment (here the shock), they still soak up residual variation in the outcome, which typically lowers standard errors. While this argument seems to suggest including a very large number of controls, to preserve degrees of freedom, it is advisable to use economic theory to select a more limited potential list of controls that are likely strong predictors of the outcome. The final set of controls and lag length can be selected based on conventional information criteria or model specification tests. In the simulations below we use the Akaike Information Criterion (AIC) for this purpose.

\end{enumerate}

\noindent The remaining two factors are specific to LP estimation. LPs that control for lagged data are referred to as \emph{lag-augmented}.

\begin{enumerate}
\setcounter{enumi}{2}

\item The third reason for including controls is that it helps robustify LP estimation. As we discussed in \cref{sec:bias_variance}, once we control for variables that are strong predictors of \emph{either} the outcome \emph{or} the impulse variable (as we should be doing anyway for the two reasons outlined above), LP estimation is relatively robust to dynamic misspecification, i.e., the omission of controls or lags with small-to-moderate predictive power. This insight applies even if the researcher has access to a credible shock measure: once lags of the strong predictors are included, we can allow for (empirically plausible) minor imperfections in the shock measure without threatening LP estimation.

\item The fourth reason for using lagged control variables is that it simplifies and robustifies the construction of LP confidence intervals. We review the reasons for this in \cref{sec:uncertainty}.

\end{enumerate}

\paragraph{Practical takeaways.}
VAR practitioners tend to select control variables and lags with an eye towards their identifying assumptions as well as predictive power, consistent with the first two factors reviewed above. The precise lag length is then often selected using the AIC or by using conventional fixed lengths such as 4 for quarterly data and 12 for monthly data, as recommended by \citet{Kilian2017}. We will review the performance of these VAR specification choices in our later simulations.

For LPs, there is instead relatively little consensus on how to select controls and lags, and so we suggest the following procedure, guided by the above considerations. In general, researchers should always control for (i) variables that are central to their identification scheme, (ii) the outcome and impulse variables, and (iii) any additional variables that strongly predict \emph{either} the outcome \emph{or} the impulse variables (or both), as suggested by past experience or economic theory. Among this set, the precise choice of lags and controls can be guided by the AIC or by other model selection criteria, proceeding as follows: first, estimate an \emph{auxiliary} VAR that includes the outcome and impulse variables and other potential controls; second, use the AIC to select the set of controls and lag length as in conventional VAR practice \citep[e.g.,][Section 2.6]{Kilian2017}; third, use the selected variables and lags as controls in subsequent LPs.\footnote{Alternatively, we could do two separate model selection exercises for a single-equation one-step-ahead forecast of the outcome and a single-equation one-step-ahead forecast of the impulse variable. We should then use the \emph{union} of the two sets of selected control variables and lags in subsequent LPs. \label{fn:double_selection}} There is no internal inconsistency in using an auxiliary VAR for selecting an LP specification: though the data-driven model selection is inevitably subject to small errors that threaten the validity of \emph{VAR-based} impulse response inference, \emph{LP-based} inference is robust to such minor model misspecification, as discussed earlier.

As a final comment on applied practice, while some LP analysts run regressions with long differences $y_{t+h}-y_{t-1}$ of the outcome variable on the left-hand side, this is actually redundant when using lagged controls. By standard OLS algebra, this outcome transformation has no impact whatsoever on the estimated LP coefficient once we control for at least one lag of the outcome, which we have argued is advisable anyway.

\subsection{Bias correction}
\label{subsec:bias_correction}

The theoretical analysis reviewed in \cref{subsec:bias_variance_analytics} focused on one key source of bias: dynamic misspecification. In the small data sets that are typical in applied macroeconomics, however, LPs and VARs are subject to another important source of bias: persistence of the data. In particular, impulse response estimates are typically biased towards displaying less persistent effects than the actual truth. For VARs, \citet{Kilian2017} recommend applying the \citet{Pope1990} bias correction to the estimated VAR coefficients to partially remove this source of bias. For LPs, \citet{Herbst2024} find that a qualitatively similar bias is present and propose a simple bias correction. Though \citet{Li2024} find that the LP bias correction increases the variance of the estimator without entirely removing bias, it is advisable to apply the correction since the entire justification for using LP over VAR prioritizes bias over variance. We demonstrate the advantages of the bias correction in practice in simulations below.

\citet{Piger2025} propose to ameliorate the small-sample LP bias by running the regression on first-differenced data and then cumulating the impulse responses to recover the level response. Whether or not to difference the data is a familiar conundrum from VAR analysis, with most textbooks preferring the levels specification (\citealp[Chapter 20.4]{Hamilton1994}; \citealp[Chapter 3.6]{Kilian2017}). The worry is that differencing the data can reduce the predictive power of the lagged controls, thus causing potentially large efficiency losses---even in large samples. In our simulations in \cref{sec:uncertainty}, bootstrap confidence intervals based on LPs that are estimated in levels and with the \citet{Herbst2024} correction robustly deliver reliable inference.\footnote{Note that, in contrast to differencing, the \citet{Herbst2024} bias correction becomes negligible (and as a result does not impair efficiency) when the sample size is large.} Hence, our tentative view is that the potential efficiency costs of the \citet{Piger2025} proposal may well outweigh the marginal benefits, though more research in this area would be welcome.

We also stress that none of the bias correction procedures discussed in this subsection deal with the bias caused by dynamic misspecification. Hence, all of our earlier points about the trade-off between LPs and VARs continue to apply to bias-corrected estimators.

\subsection{Shrinkage}

Disappointed by the often jagged-looking impulse response functions that are produced by least-squares LPs and VARs, some researchers have turned to variants of these methods that employ shrinkage---i.e., nudging the least-squares estimate in a direction that is regarded as \emph{a priori} plausible. The penalized LP estimator of \citet{Barnichon2019} smooths out the estimated impulse responses across horizons by adding a term to the OLS objective function that penalizes deviations from a quadratic impulse response function, with the overall degree of shrinkage determined by cross-validation. Similarly, Bayesian VARs shrink unrestricted estimates towards simpler dynamics, such as independent random walks or white noise processes \citep{Doan1984,Todd1984,Litterman1986,Giannone2015}. Finally, model averaging or selection techniques use data-dependent rules to partially or fully shade the LP estimator toward a short-lag VAR estimate \citep{Ferreira2023,Gonzalez2025}.

While the attraction of these shrinkage estimators is that they reduce the variance of the jagged least-squares estimators, they will do so explicitly by introducing additional bias. Whether the added bias is actually worthwhile depends both on the DGP and on the researcher's objective function, as we discuss in the next two sections.

\section{The bias-variance trade-off in practice}
\label{sec:biasvar_practice}

This section uses empirically calibrated simulations to quantify the bias-variance trade-off between various implementations of LPs and VARs. The simulation evidence complements our theoretical treatment, revealing what shape the bias-variance trade-off is likely to take in practice, and so laying the groundwork for our later practical recommendations.

\subsection{The simulation experiment}

Prior simulation studies comparing the performance of LP and VAR methods include \citet{Kilian2011}, \citet{Bruns2021}, and \citet{Li2024}; we here build on and extend the latter study. Just like those authors, we consider a large menu of possible DGPs, all estimated to mimic as closely as possible the properties of the universe of U.S. macroeconomic data. We provide a high-level overview of the DGPs here, with implementation details relegated to the original work of \citeauthor{Li2024} as well as to \cref{app:simulations_setup}.

\paragraph{Data generating processes.}
Our simulations are based on a dynamic factor model (DFM) fitted to a large number of standard U.S. macroeconomic time series. We use the data set of \citet{Stock2016}, which consists of 207 quarterly U.S. time series spanning a range of variable categories, from quantities to prices. We then fit two separate DFMs to this data set: a non-stationary variant that allows for cointegrating relationships among the factors, as well as a stationary one for which all data series have been pre-transformed to ensure stationarity. Differently from \citeauthor{Li2024}, we allow for conditional heteroskedasticity of the shocks in the equations for both the factors as well as the idiosyncratic disturbances. The heteroskedasticity is modeled through ARCH processes with empirically estimated parameters. Overall, the resulting DFMs imply complex and varied dynamics for macroeconomic time series of the sort encountered in applied practice.

Given these DFMs, we proceed to construct a diverse array of lower-dimensional DGPs by considering hundreds of different subsets of time series, with variable selection closely emulating applied practice. Specifically, each DGP consists of a set of five observable time series, selected at random from those of the DFM's series that are most commonly used in applied practice. We then consider a researcher who observes data of those particular selected series, and identifies monetary and fiscal policy shocks through either a recursive ordering of these observable time series, or by also additionally measuring the policy shock, as in our discussion of identification in \cref{sec:identification}. For the monetary policy DGPs we restrict the vector of observables to always contain the federal funds rate, while for fiscal policy DGPs we always include government spending; for recursive identification these two policy variables are ordered last and first, respectively. The researcher is then interested in the response of one of the other four variables to the identified shock. She estimates this response using several different estimation strategies, to be discussed in detail below. Given that the true DFM is known to us, we can through simulations compute estimator biases, variances, and mean squared errors (for this section) as well as confidence interval properties (for the uncertainty assessments in the next section).

We stress that the DFMs and the resulting DGPs  that we construct do not admit finite-order VAR representations, yielding a non-trivial bias-variance trade-off between LP and VAR estimators. The latent nature of the macroeconomic factors and the idiosyncratic disturbances induce moving average dynamics in the processes for the observed time series, consistent with our discussion of dynamic misspecification in \cref{subsec:bias_variance_analytics}.

Our overall objective is to provide an empirically grounded assessment of which estimators are likely to perform well on average in settings typically encountered in applied work, and thus can serve as attractive default procedures. While we cannot claim that our empirically calibrated DFMs capture all aspects of the real world, we contend that a minimum condition for an econometric procedure to be useful for applied work is that it should at least perform well on the kinds of DGPs that we consider. The empirically grounded DGPs will furthermore showcase the quantitative bite of the general theoretical considerations reviewed in \cref{sec:bias_variance}.

\paragraph{Estimators.}
We consider several variants of the LP and VAR estimators introduced in \cref{sec:identification}---variants that differ in the selection of the lag length $p$, the choice of control variables $w_t$, and whether we apply the shrinkage techniques discussed in \cref{sec:lpvar_details}. All estimators include an intercept. We only provide a brief overview here, with further technical details in \cref{app:lp_var_details}.

\begin{enumerate}[1.]

\item {\bf LP.} The researcher estimates the LP \eqref{eq:lp_pop} to implement her observed shock or recursive identifying assumptions. We consider several different LP variants.

\begin{itemize}

\item {\it Lag length.} We either fix the lag length at conventional values, or estimate it using the Akaike Information Criterion (AIC) following the procedure that was outlined in \cref{subsec:lags_controls}.

\item {\it Control variables.} The researcher observes a list of macroeconomic aggregates. Under recursive shock identification, all of those observables need to be included as controls. Under observed shock identification, we consider two specifications: one that controls for lags of all of the observed variables, and a smaller one that controls for only lags of the measured shock and the outcome of interest.

\item {\it Shrinkage.} We consider the penalized LP estimator of \citet{Barnichon2019}, with the type and amount of shrinkage chosen exactly as recommended by those authors, lag length fixed at conventional values, and the full set of controls.

\end{itemize}

The non-shrinkage LP estimators apply the bias correction of \citet{Herbst2024}, as discussed in \cref{subsec:bias_correction}.

\item {\bf VAR.} The researcher estimates the VAR \eqref{eq:var_pop} to implement her identifying assumptions. We also consider several different VAR variants.

\begin{itemize}

\item {\it Lag length.} The lag length is either fixed at conventional values or estimated using the AIC for the reduced-form VAR, consistent with our discussion in \cref{subsec:lags_controls}.

\item {\it Control variables.} Just as for LP, we include the full set of observables as controls for recursive identification, while for observed shock identification we include either the full set or estimate a bivariate (``internal instrument'') system in shock and outcome.

\item {\it Shrinkage.} Our implementation of the Bayesian VAR adopts the rich prior specification of \citet{Giannone2015}, with prior hyperparameters selected by maximizing the marginal likelihood, lag length fixed at conventional values, and including the full set of controls.

\end{itemize}

The non-Bayesian VAR estimators apply the \citet{Pope1990} analytical correction for biases caused by persistent data, consistent with the recommendation of \citet{Kilian2017}, and as discussed in \cref{subsec:bias_correction}.

\end{enumerate}

\subsection{Results}
\label{subsec:biasvariance_simul}

In this section we present simulation results for 200 stationary and 200 non-stationary DGPs, for observed shock identification and averaging across monetary and fiscal shocks, with 100 DGPs for each. Results separately by the type of shock and for recursive identification are broadly similar, and relegated to \cref{app:fiscal_monetary,app:recursive}. We approximate population biases and variances by averaging across 1,000 Monte Carlo simulations per DGP.

\begin{figure}[tp!]
\centering
{\textsc{Observed shock, stationary DGPs}} \\ 
\vspace{\baselineskip}
\begin{subfigure}[c]{0.49\textwidth}
\includegraphics[width=\textwidth]{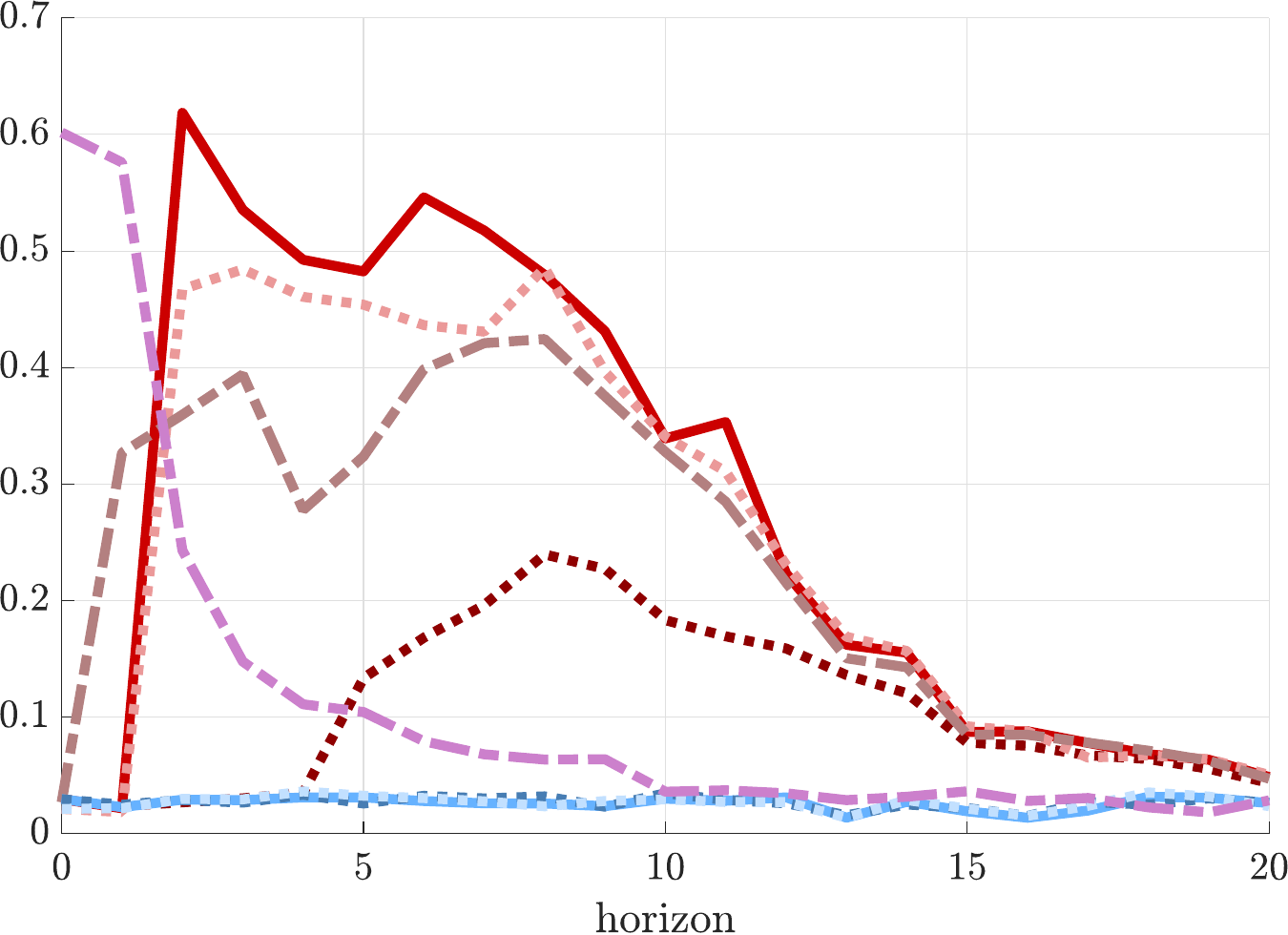}\vspace{0.2cm}
\caption*{\textsc{Bias}} \vspace{0.6cm}
\end{subfigure}
\begin{subfigure}[c]{0.49\textwidth}
\includegraphics[width=\textwidth]{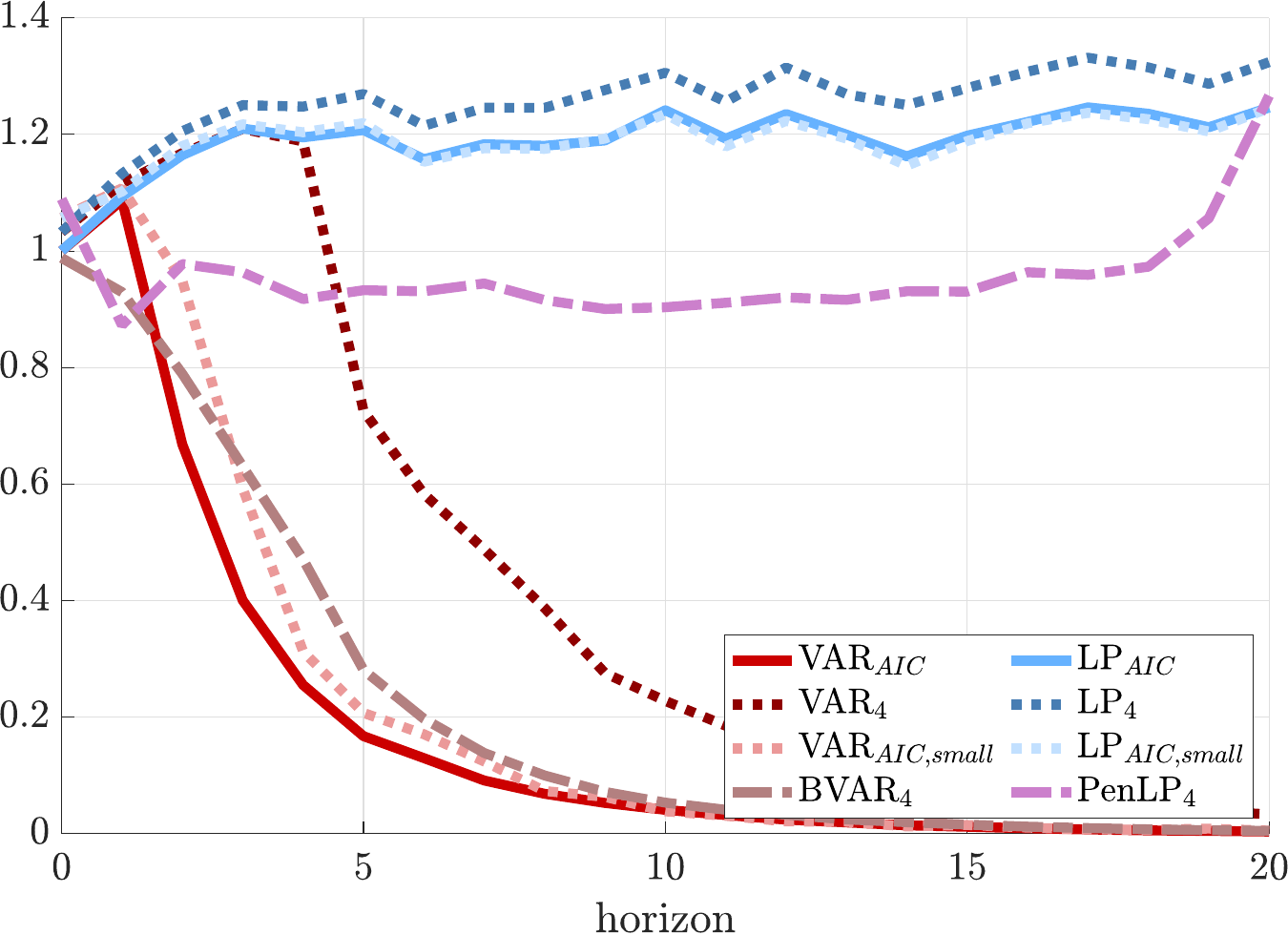}\vspace{0.2cm}
\caption*{\textsc{Standard deviation}} \vspace{0.6cm}
\end{subfigure}
\centering
{\textsc{Observed shock, non-stationary DGPs}} \\ 
\vspace{\baselineskip}
\begin{subfigure}[c]{0.49\textwidth}
\includegraphics[width=\textwidth]{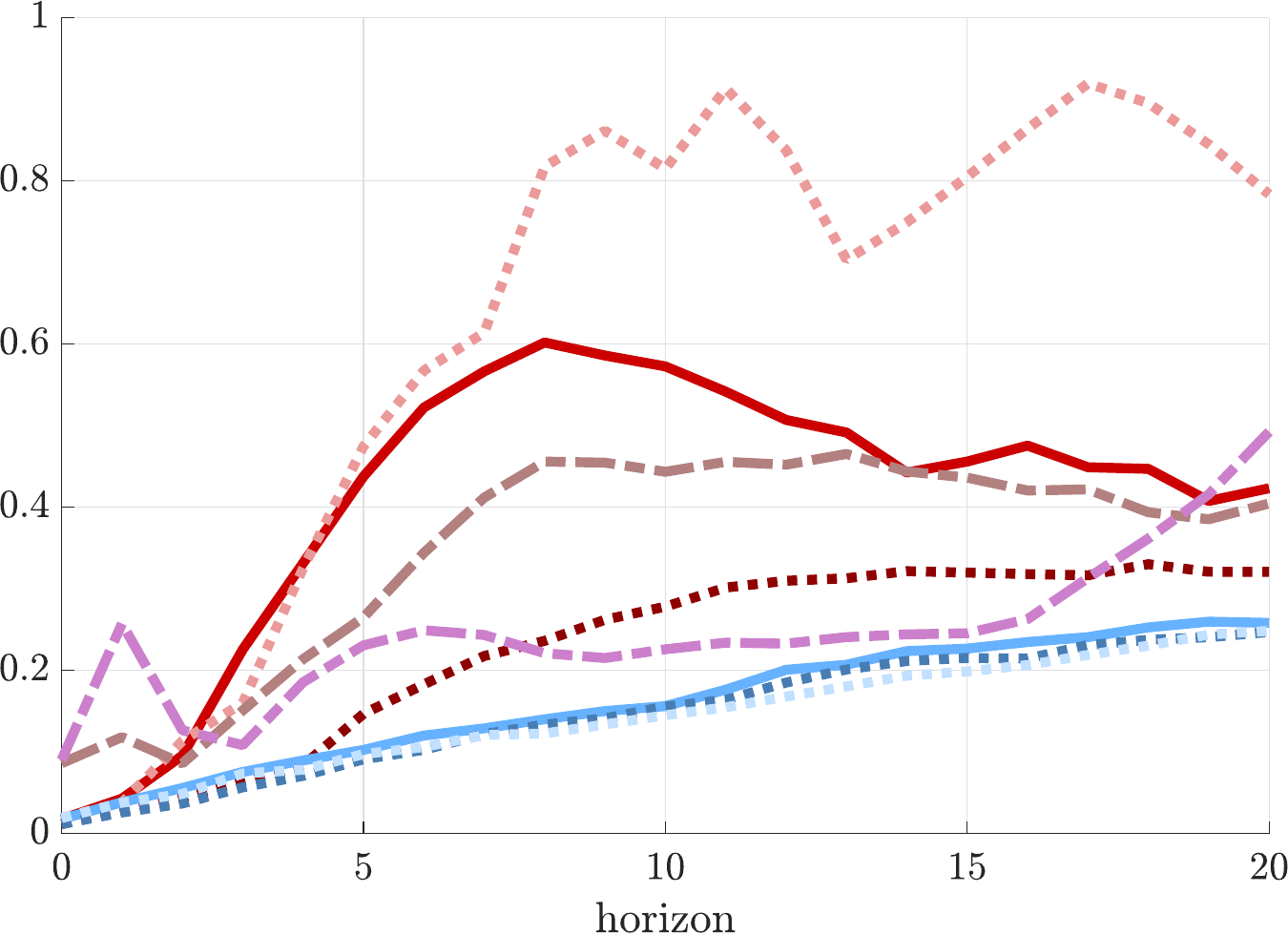}\vspace{0.2cm}
\caption*{\textsc{Bias}} \vspace{0.2cm}
\end{subfigure}
\begin{subfigure}[c]{0.49\textwidth}
\includegraphics[width=\textwidth]{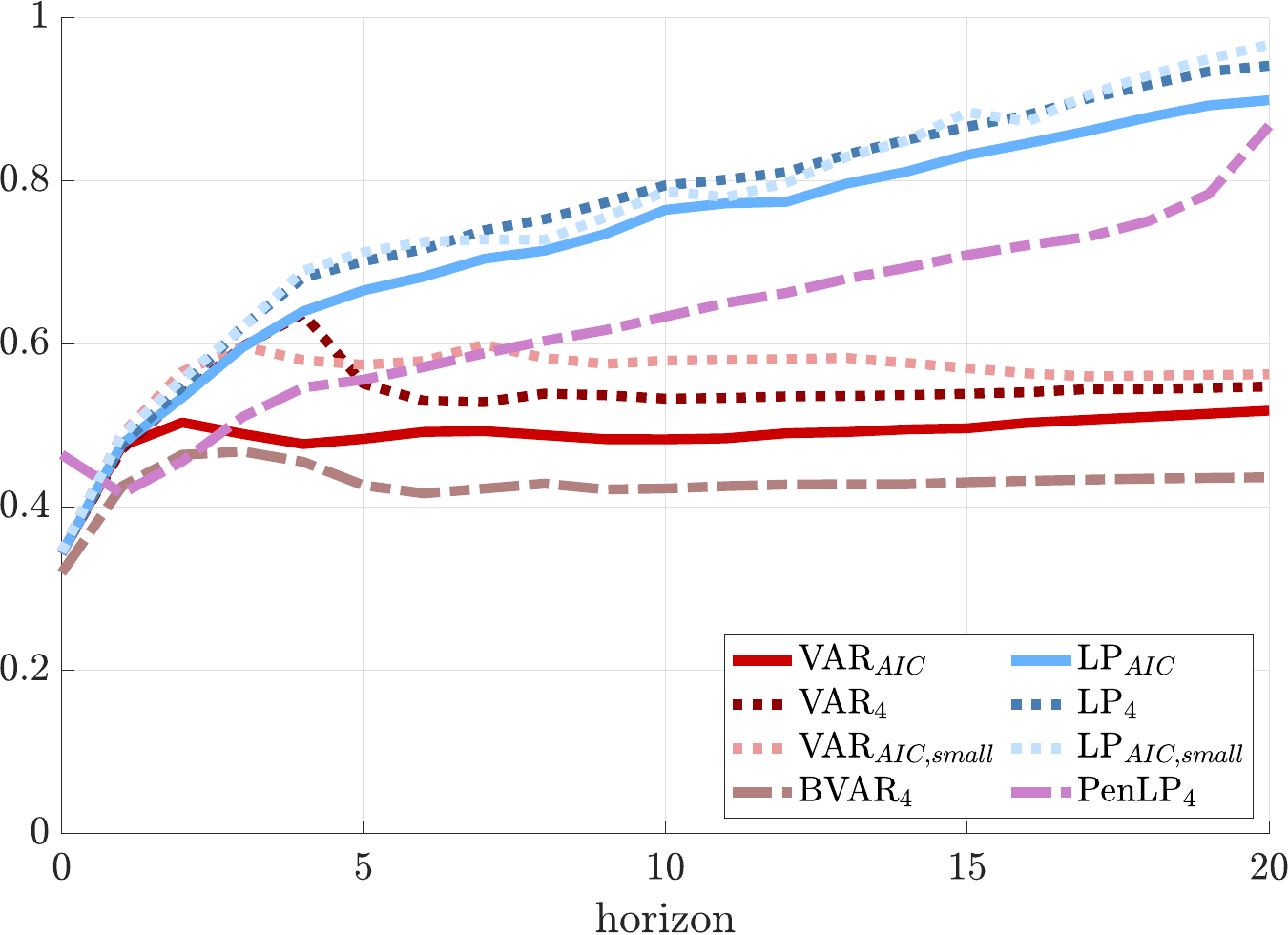}\vspace{0.2cm}
\caption*{\textsc{Standard deviation}} \vspace{0.2cm}
\end{subfigure}
\caption{Median (across DGPs) of absolute bias $ |\mathbb{E} [ \hat{\theta}_h-\theta_h ] |$ (left panels) and standard deviation $\sqrt{\var(\hat{\theta}_h)}$ (right panels) of the different estimation procedures, relative to $\sqrt{\frac{1}{21}\sum_{h=0}^{20}\theta_h^2}$. The subscript ``AIC'' indicates lag length selection via the AIC, ``4'' indicates four lags, ``small'' indicates a small system containing only shock and outcome of interest, ``BVAR'' indicates Bayesian VAR, and ``PenLP'' indicates penalized LP.}
\label{fig:biasstd_obsshock}
\end{figure}

\paragraph{Bias-variance trade-off.}
\cref{fig:biasstd_obsshock} reveals that, in applied practice, researchers do indeed face a stark bias-variance trade-off between LPs, VARs, and intermediate shrinkage techniques, consistent with our earlier theoretical discussion. The left panels of the figure show the median absolute bias (across all the monetary and fiscal policy shock DGPs) for the stationary (top) as well as the non-stationary (bottom) encompassing DFM, while the right panels show the median standard deviation, in both cases normalized by the overall scale of the true impulse response. Red lines correspond to VAR estimators, and blue lines indicate LP estimators, with the various line styles corresponding to several different options for the selection of lag length, controls, and shrinkage.\footnote{We only report results for bias-corrected LPs and VARs, consistent with our discussion in \cref{subsec:bias_correction}. Bias correction barely matters in the stationary DGPs, but LP biases would be materially larger in non-stationary DGPs in the absence of bias correction, particularly at medium and long horizons.} While all least-squares LP estimators have relatively low bias across horizons, the VAR estimators tend to suffer from substantial bias at intermediate horizons, as well as at long horizons in the more persistent DGPs. Finally, and as expected, shrinkage---here in the form of either penalized LPs or Bayesian VARs---reduces variance at the cost of higher bias.

While LPs are relatively insensitive to dynamic specification, the bias-variance properties of VAR estimators are shaped decisively by the estimation lag length, the horizon of interest, and the choice of controls. The AIC typically selects a very short lag length in our DGPs (the mean lag length selected equals $p = 1.88$), and so we see a sharp bias-variance trade-off already at short impulse horizons. Manually increasing the VAR lag length to $p = 4$ (dark red, dashed) aligns LPs and VARs up to horizon $h = 4$, again consistent with the theory reviewed earlier. The performance of LPs is, on the other hand, virtually unaffected by the precise number of estimation lags (in solid blue vs.\ dashed dark blue).\footnote{\cref{app:recursive} shows that the number of lags plays a more important role under recursive identification, since then the lags also matter for spanning the shock of interest (recall \cref{subsec:lags_controls}).} Furthermore, in the stationary DGPs, both the VAR bias and standard deviation are small at long horizons, since all impulse responses converge to zero relatively fast. In practice, however, the speed of this convergence is uncertain; in particular, in the non-stationary DGPs (displayed in the bottom panel), the bias-variance trade-off remains stark even at much longer horizons. Finally, comparing the solid and light-dashed lines, we see that the effect of including either few or many controls is negligible for LPs, but important for VARs.

\begin{figure}[tp!]
\centering
{\textsc{Observed shock: MSE}} \\ 
\vspace{1\baselineskip}
\begin{subfigure}[c]{0.49\textwidth}
\includegraphics[width=\textwidth]{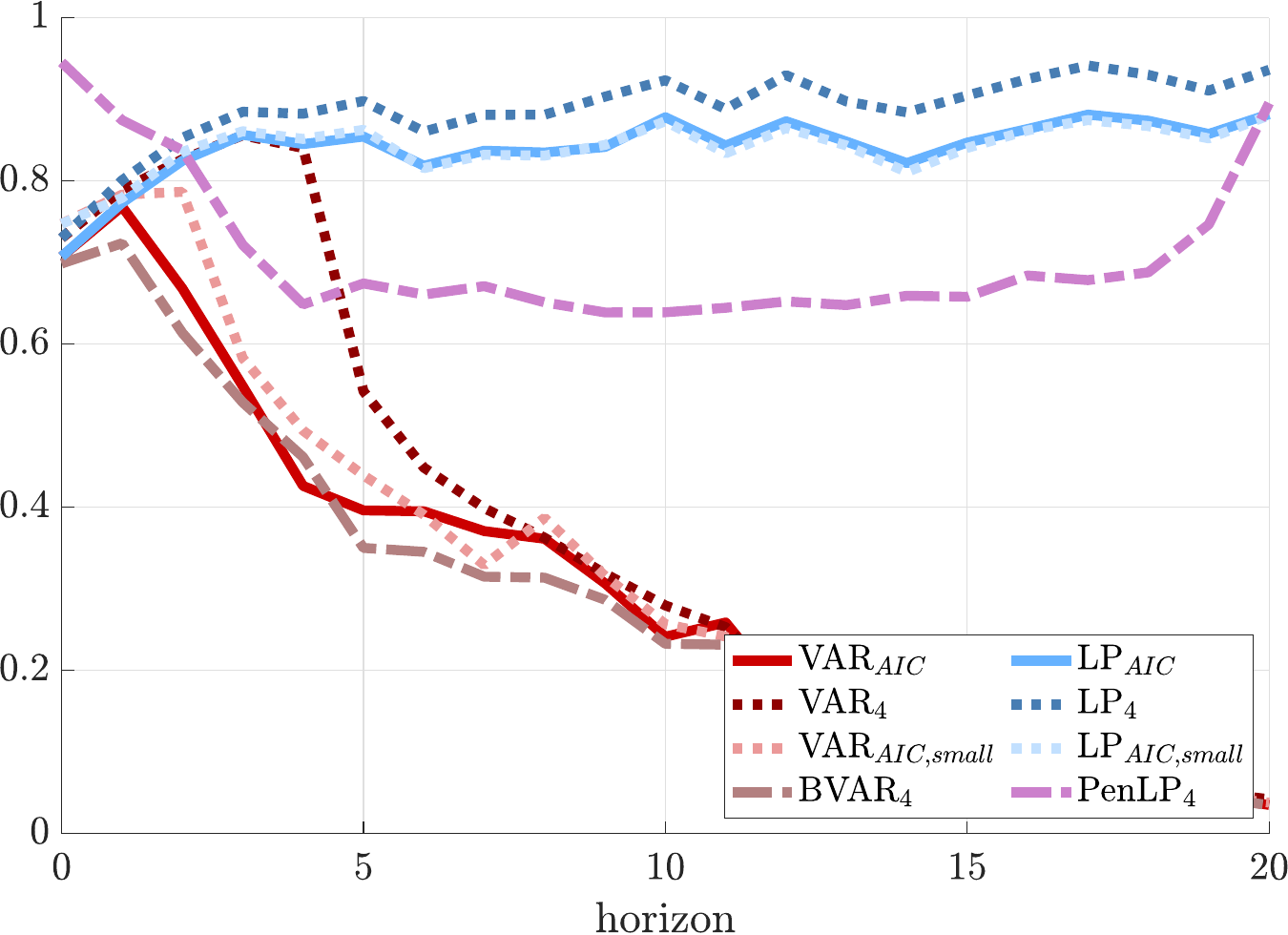}\vspace{0.2cm}
\caption*{\textsc{Stationary DGPs}} \vspace{0.4cm}
\end{subfigure}
\begin{subfigure}[c]{0.49\textwidth}
\includegraphics[width=\textwidth]{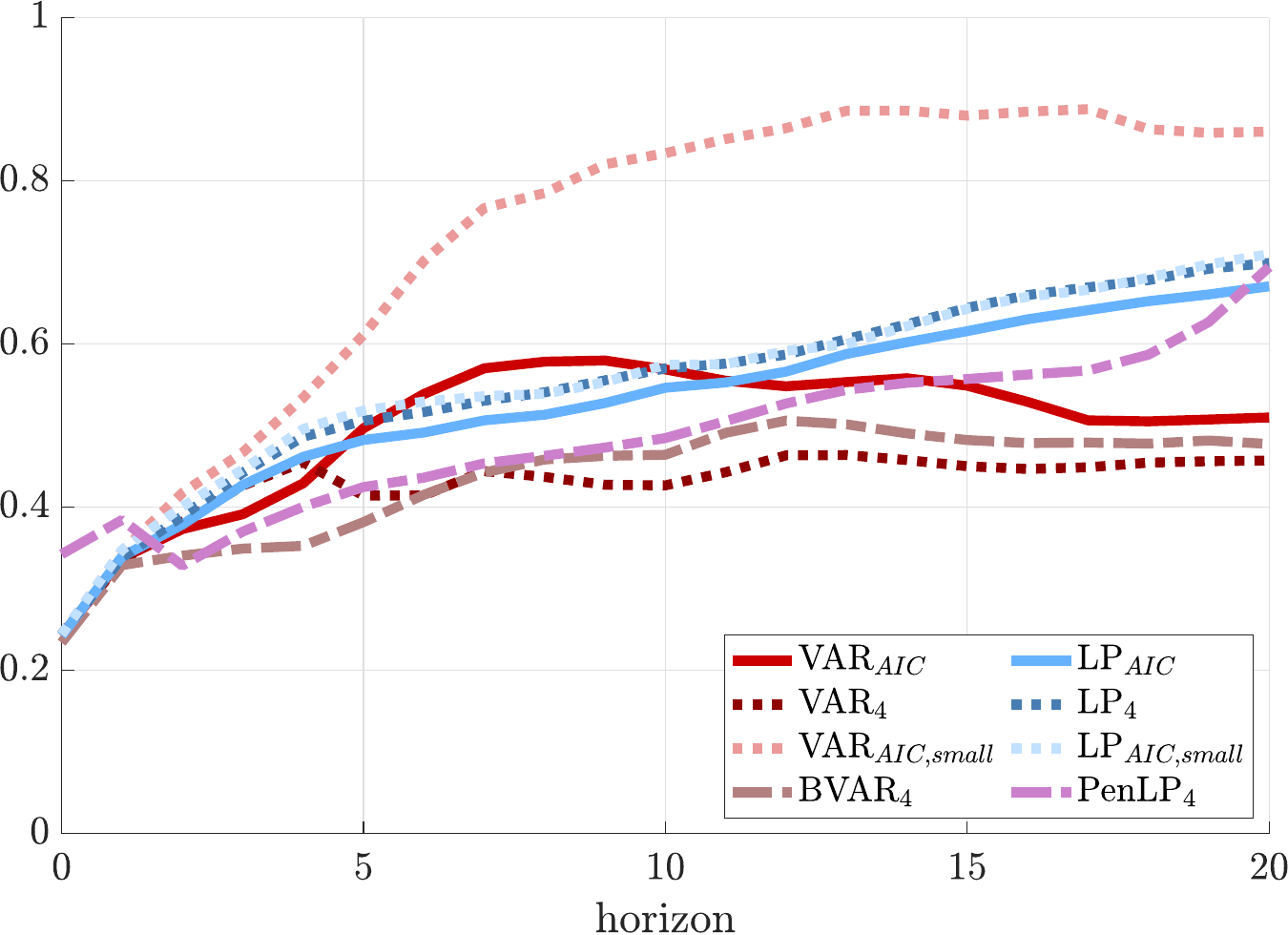}\vspace{0.2cm}
\caption*{\textsc{Non-Stationary DGPs}} \vspace{0.4cm}
\end{subfigure}
\caption{Median (across DGPs) of mean squared error $\MSE(\hat{\theta}_h)$ of the different estimation procedures, relative to $\sqrt{\frac{1}{21}\sum_{h=0}^{20}\theta_h^2}$, for the stationary DGPs (left panel) and the non-stationary DGPs (right panel).}
\label{fig:mse_obsshock}
\end{figure}

\paragraph{Mean squared error.}
We now further assess the quantitative nature of the bias-variance trade-off using the familiar mean squared error (MSE) criterion. For a given estimator $\hat{\theta}_h$ of the true impulse response $\theta_h$, the MSE is defined as
\begin{equation*}
\MSE(\hat{\theta}_h) = E\left[\left(\hat{\theta}_h-\theta_h \right)^2 \right] = \underbrace{\left(\mathbb{E} \left [ \hat{\theta}_h - \theta_h \right ]\right)^2}_{\bias(\hat{\theta}_h)^2} + \var(\hat{\theta}_h).
\end{equation*}
In words, MSE weights equally an estimator's (squared) bias and variance. \cref{fig:mse_obsshock} plots the MSE of the estimators, again separately for stationary (left) and non-stationary (right) DGPs. We see that MSE is typically minimized by a VAR method, either Bayesian or least-squares. Thus, while we know theoretically that there exist \emph{some} DGPs close to finite-order VARs for which LPs outperform VARs in terms of MSE, VARs are favored in the \emph{typical} DGP.\footnote{According to the worst-case bias result \eqref{eq:bias_bound} in \cref{subsec:bias_variance_analytics}, even a minor amount of misspecification $\mathcal{M} \geq 1/T$ could suffice for VAR MSE to exceed LP MSE \citep[see][Section 4.1]{Montiel2024}. In the DGPs that we consider, the \emph{magnitude} of misspecification is indeed quite material (e.g., for the stationary DFM, the average $\sqrt{T \times \mathcal{M}}$ across DGPs is $2.64$ for $p = 4$ VAR lags). However, it turns out that the \emph{type} of misspecification is not close to the worst case \emph{on average}, causing the VAR MSE to typically be smaller. We thank Isaiah Andrews for raising this point. \label{fn:avg_misspecification}} This finding is a consequence of the fact that, across our DGPs, the bias reduction of LP typically comes at more than a one-to-one cost in terms of the standard deviation.

\begin{lesson}
\label{lesson_biasvar_simul}

Empirically calibrated simulations reveal that, in practice, there is a sharp bias-variance trade-off between LPs and VARs at intermediate and long horizons. LPs attain their low bias at significant variance cost. If the researcher's objective is to minimize mean squared error, then she will \emph{typically} be unwilling to incur that variance cost.

\end{lesson}

This lesson is the main takeaway of \citet{Li2024}, and it is also consistent with the conclusions of \citet{Marcellino2006} in a forecasting context. Some researchers may, however, be more concerned with minimizing bias than minimizing variance, in which case plain LPs could still be preferable over VARs (and the shrinkage estimators). In the next section we will argue that if the goal is not merely to construct a point estimate but also to accurately convey the uncertainty surrounding the estimate, then we are  \emph{endogenously} forced to heavily prioritize bias.

\section{Uncertainty assessments}
\label{sec:uncertainty}

Applied macroeconomists not only report point estimates of dynamic causal effects, they also want to quantify the statistical uncertainty of those estimates. This is an important task, since the standard errors are often of roughly the same magnitude as the estimates themselves. The accuracy of uncertainty assessments is conventionally evaluated by the \emph{coverage} of the implied confidence interval: the probability that the reported interval covers the true impulse response should be at least 90\% (say), at every horizon and regardless of the shape of the true impulse response function.

This section argues that LPs---or VARs with very long lag lengths---are the only known procedures that can robustly achieve satisfactory coverage of confidence intervals in practice. The intuition is that even quite small amounts of bias (relative to the standard error) can cause severe coverage distortions, and as we have seen, only LPs (or equivalently VARs with very long lags) robustly achieve low bias across a range of empirically relevant DGPs. In other words, while short-lag VARs deliver point estimates that are accurate most of the time, only LPs deliver confidence intervals that are valid (almost) all the time.

\subsection{Confidence intervals for impulse responses}
\label{subsec:lp_ci}

We begin by reviewing methods for computing standard errors for LP estimation of impulse responses. We will not review VAR-based inference methods, since these are already covered in textbooks such as \citet{Kilian2017}.

For LP impulse responses, heteroskedasticity-robust standard errors suffice for accurate uncertainty assessments. In the baseline LP regression \eqref{eq:lp_pop}, the residual $\xi_{h,t}$ is typically serially correlated because it is a multi-step forecast error, which would suggest the use of Heteroskedasticity and Autocorrelation Consistent (HAC) standard errors, such as Newey-West. Fortunately, however, \citet{MontielOlea2021} show that HAC corrections are unnecessary under a weak assumption on the shocks, and conventional heteroskedasticity-robust standard errors for OLS suffice. The reason is that, while the LP regression residual $\xi_{h,t}$ is indeed serially correlated, what matters for the distribution of the LP estimator is the \emph{product} of the residual and the residualized shock $\tilde{x}_t$ defined in \eqref{eq:xtilde}. This product is serially uncorrelated (though typically heteroskedastic) under the natural assumption that the shocks are unpredictable (i.e., conditionally mean independent) from their past and future values. Even if this assumption is slightly violated, it is still likely that the well-documented practical challenges of HAC estimation \citep{Lazarus2018,Herbst2024} will make the conventional heteroskedasticity-robust standard errors a better choice for applied work.

Given a standard error $\hat{\tau}_{h,\text{LP}}$ for the LP impulse response estimate $\hat{\theta}_h^\text{LP}$, a level-$(1-a)$ confidence interval can be obtained with the usual formula:
\begin{equation*}
\hat{\theta}_h^\text{LP} \pm \hat{\tau}_{h,\text{LP}} z_{1-a/2},
\end{equation*}
where $z_{1-a/2}$ is the $(1-a/2)$ quantile of the standard normal distribution (e.g., $z_{1-a/2} \approx 1.64$ for a $1-a=90\%$ confidence interval). \citet{MontielOlea2021} find that a bootstrap confidence interval can further improve coverage in small samples. However, the bootstrap they consider is only valid for reduced-form impulse responses, not structural impulse responses. Our simulation study in \cref{subsec:inference_simul} suggests that confidence intervals obtained by bootstrapping LP estimates using the residual block bootstrap of \citet{Brueggemann2016} have accurate coverage for structural impulse responses.

\begin{lesson}
When computing confidence intervals for LP impulse responses, heteroskedasticity-robust standard errors typically perform as well or better than more complicated HAC corrections. Simulation evidence suggests that further improvements in finite-sample performance can be obtained through a bootstrap algorithm, such as the one in \cref{app:lp_var_details}.
\end{lesson}

The next sections compare---first theoretically, then through simulations---the quality of uncertainty assessments based on either LP or VAR inference.

\subsection{Bias is very costly for coverage}
\label{subsec:bias_coverage}

In this section we establish theoretically that conventional VAR confidence intervals can exhibit severe coverage distortions even under relatively small amounts of dynamic misspecification, while LP intervals are instead much more robust. Our analysis here builds on and extends our review of the bias-variance trade-off in \cref{subsec:bias_variance_analytics}.

\paragraph{How bias affects coverage.}
The approximate VAR sampling distribution in \eqref{eq:large_sample_distr} together with a straightforward calculation reveals that the coverage probability of the conventional VAR confidence interval is a decreasing function of the bias/standard-error ratio $|b_h(p)|/\tau_{h,\text{VAR}}(p)$; specifically, it equals
\begin{equation}
P(|Z| \leq z_{1-a/2}), \quad \text{where $Z \sim N(|b_h(p)|/\tau_{h,\text{VAR}}(p),1)$}. \label{eq:var_coverage}
\end{equation}
The LP confidence interval, on the other hand, robustly achieves the nominal coverage level, by virtue of LP's zero (asymptotic) bias.

\begin{figure}[tp]
\centering
\textsc{Coverage of VAR confidence intervals} \\ 
\vspace{1\baselineskip}
\includegraphics[width=0.6\linewidth]{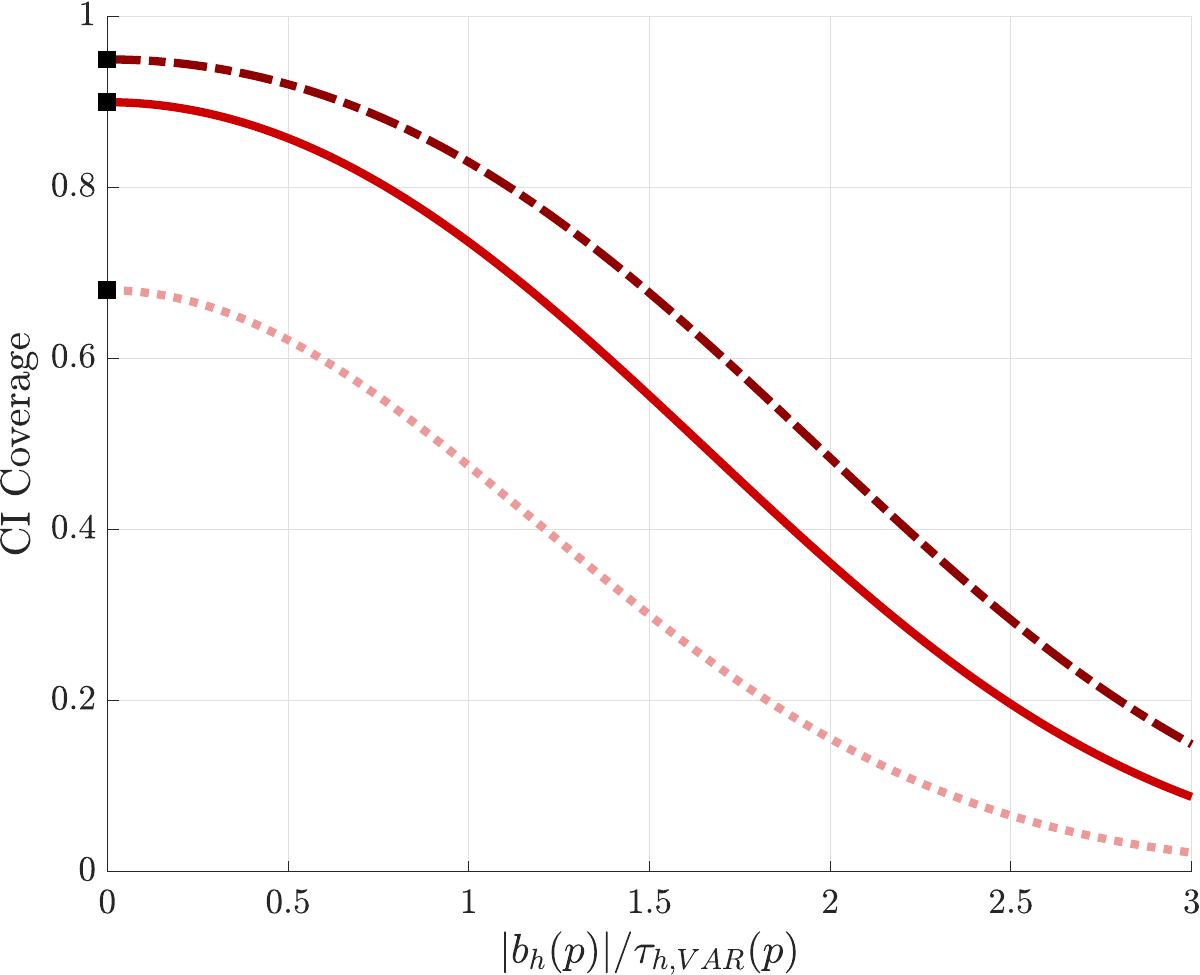}
\caption{Asymptotic coverage of VAR confidence interval as a function of relative VAR bias $|b_h(p)|/ \tau_{h,\text{VAR}}(p)$, for target coverage levels of 68\% (dotted), 90\% (solid), as well as 95\% (dashed-dotted). The black squares indicate the target coverage level.}
\label{fig:VAR_coverage}
\end{figure}

\cref{fig:VAR_coverage} plots the coverage probability of the VAR confidence interval as a function of the (scaled) VAR bias and for a range of target coverage levels. We see that even a moderate ratio of bias to standard error yields large coverage distortions. In \cref{subsec:bias_variance_analytics} we gave a numerical example in which a small amount of VAR misspecification caused the bias to be around 2.29 times the standard error; this ratio would cause a putative 90\% confidence interval to cover the true impulse response with probability less than 30\%! The right panel of \cref{fig:lp_vs_var_ramey} indeed suggests that the bias/standard-error ratio of VARs exceeds 2 at long horizons in many applications.\footnote{The scaled difference $(\hat{\theta}_h^\text{LP}-\hat{\theta}_h^\text{VAR})/\tau_{h,\text{VAR}}(p)$ of LP and VAR estimates is an unbiased estimate of $b_h(p)/\tau_{h,\text{VAR}}(p)$. An important caveat is that \cref{fig:lp_vs_var_ramey} reports the absolute value of the scaled difference, which is an overestimate of the absolute scaled bias.} \cref{fig:VAR_coverage} thus implies that a researcher who is interested in guaranteeing that the coverage of a reported confidence interval is not too low relative to the target coverage must \emph{necessarily} prioritize bias over variance.

Because bias is so costly for coverage, LP confidence intervals---or equivalently VAR confidence intervals with a very large estimation lag length---are necessarily preferred over short-lag VARs when it comes to uncertainty assessments. Recall that, in the face of minor amounts of misspecification, we can only guarantee a low bias/standard-error ratio for VARs when we use very many lags (usually many more lags than indicated by conventional model selection or evaluation procedures). In practice, it thus only appears to be reasonable to trust VAR confidence intervals if they are approximately as wide as corresponding LP intervals.

\begin{lesson}
The coverage of conventional VAR confidence intervals is highly sensitive to dynamic misspecification, unlike LP intervals. To robustify VAR intervals, one must control for so many lags that the intervals become approximately the same as the LP intervals.
\end{lesson}

We furthermore stress that one does not need to believe that the (short-lag) VAR bias is large in \emph{most} applications in order to prefer LP over VAR confidence intervals. In conventional statistical and econometric practice, we seek to control coverage \emph{uniformly} across a wide range of empirically relevant DGPs, and not merely for the ``typical'' DGP. In other words, our uncertainty assessment should reliably indicate that uncertainty is high \emph{whenever} that is in fact the case. Hence, the mere fact that VARs can \emph{sometimes} be badly biased, as shown theoretically in \cref{subsec:bias_variance_analytics} and then practically in \cref{sec:biasvar_practice}, would militate against short-lag VAR confidence intervals. Our simulations in \cref{subsec:inference_simul} will illustrate these theoretical observations.

\paragraph{MSE vs.\ coverage.}
To reiterate, the optimal procedure when it comes to producing confidence intervals with correct coverage is not the same as the MSE-minimizing procedure. Based on the approximate distributions \eqref{eq:large_sample_distr} of LP and VAR, a simple calculation shows that VARs are preferred to LP on MSE grounds if and only if
\begin{equation*}
|b_h(p)| / \tau_{h,\text{VAR}}(p) \leq \sqrt{ \tau_{h,\text{LP}}^2/\tau_{h,\text{VAR}}^2(p) - 1 }.
\end{equation*}
Once again making reference to the right panel of \cref{fig:lp_vs_var_ramey}, and focusing on long horizons, we may expect the left-hand side of the above inequality to be close to 2 in many applications. The left panel of the same figure shows that the median value of $\tau_{h,\text{LP}}/\tau_{h,\text{VAR}}(p)$ can be close to 0.4 in practice, corresponding to a value of $\sqrt{5.25} \approx 2.29$ for the right-hand side of the above inequality. Thus, in ``typical'' applications, VAR point estimators are preferable from the MSE perspective, yet the associated VAR confidence interval can have poor coverage.

\paragraph{Shrinkage, model selection, and Bayesian inference.}
If we seek confidence intervals that have correct coverage regardless of what the true impulse response function looks like, then it is likely impossible to improve much upon the plain LP confidence interval, at least in large samples. One might hope that shrinkage, model averaging, or model selection techniques could be used to develop shorter confidence intervals that ``adapt'' to the nature of the DGP; that is, decide in a data-dependent manner to what extent we should report wide LP intervals or some narrower interval that imposes smoothness or parametric structure. However, when applied to the theoretical framework of \cref{sec:bias_variance}, mathematical impossibility results from microeconometrics \citep{Pratt1961,Armstrong2018} strongly suggest that the potential for such adaptation is very modest. In other words, we can only get away with reporting (meaningfully) narrower confidence intervals than LP if we give up on the usual notion of coverage.\footnote{\citet{Freyaldenhoven2024} propose one such way to change the researcher objective.} Indeed, to our knowledge, none of the existing confidence intervals based on shrinkage procedures for LPs or VARs have frequentist coverage guarantees comparable to that of the plain LP interval, and we will see below that some of them indeed have poor coverage in simulations.

A Bayesian who is completely confident in their VAR model and prior can of course ignore LP evidence; however, any Bayesian who is worried about slight model misspecification is forced to use LP as a diagnostic in the same way as a frequentist. This is because, in large samples, the Bayesian VAR posterior distribution behaves just like the sampling distribution of a frequentist VAR estimator, and we have already seen that the latter is highly sensitive to small amounts of dynamic misspecification.

\subsection{Challenges at long horizons and with persistent data}
\label{subsec:longrun_ci}

While the previous subsection discussed the fragility of VAR inference under mild misspecification, LP confidence intervals can in fact also have more accurate coverage than conventional VAR confidence intervals \emph{even if the VAR model is correctly specified}. We review the intuition here and refer to \cite{MontielOlea2021} for technical details.

\citeauthor{MontielOlea2021} show that, assuming a correctly specified VAR model, LP confidence intervals control coverage more robustly at long horizons and across a wider range of persistence of the underlying DGP than VAR intervals. Intuitively, at long horizons $h$, the VAR impulse response is a highly nonlinear transformation of the estimated parameters (e.g., recall the exponential formula $\hat{\rho}^h$ in the AR(1) case). This spells trouble for traditional VAR inference procedures because they ultimately rely on a linearization argument for asymptotic validity. By contrast, LP simply relies on linear regressions. Moreover, in DGPs with (near-)unit roots, it is well known that autoregressive coefficient estimates can have non-normal distributions, which dramatically complicates the calculation of appropriate critical values for impulse responses. LP, however, is equivalent to a projection of the outcome on the residualized shock $\tilde{x}_t$ in \eqref{eq:xtilde}, and the latter is robustly stationary (provided we employ lag augmentation as recommended in \cref{sec:lpvar_details}), so that the LP coefficient will typically still have a normal distribution even if the data has stochastic trends.

\begin{lesson}
Lag-augmented LP confidence intervals are more robust than conventional VAR confidence intervals to the length of the impulse response horizon and the persistence of the data, even if the VAR model is correctly specified.
\end{lesson}

\subsection{Simulation evidence}
\label{subsec:inference_simul}

We now complement the theoretical insights of the previous two subsections with simulation results. Our simulations are again based on the large menu of DGPs described in \cref{sec:biasvar_practice}.

\paragraph{Confidence intervals.}
We report the coverage of confidence intervals constructed from several variants of LP and VAR estimators, with a target confidence level of 90\%. Further implementation details are provided in \cref{app:lp_var_details}.

\begin{enumerate}[1.]

\item {\bf LP.} We consider three approaches to constructing LP confidence intervals.

\begin{itemize}

\item {\it Analytical.} This analytical confidence interval is constructed using conventional heteroskedasticity-robust standard errors for OLS, deliberately ignoring HAC corrections, as discussed in \cref{subsec:lp_ci}.

\item {\it Bootstrap.} We bootstrap LP estimates by generating samples from the VAR residual block bootstrap of \citet{Brueggemann2016}, as discussed in \cref{subsec:lp_ci}. We report the percentile-t bootstrap confidence interval. 

\item {\it Shrinkage.} We report a confidence interval centered at the penalized LP estimator of \citet{Barnichon2019}, with the heuristic standard errors suggested by those authors and normal critical value.

\end{itemize}

\item {\bf VAR.} We also consider three approaches for VAR confidence intervals.

\begin{itemize}

\item {\it Analytical.} We use textbook formulae to compute delta method standard errors and associated confidence intervals for the VAR estimators.

\item {\it Bootstrap.} We use the residual block bootstrap of \citet{Brueggemann2016}. Following the recommendation of \citet{Inoue2020}, we report the Efron bootstrap confidence interval.

\item {\it Shrinkage.} For a Bayesian alternative, we report the equal-tailed posterior credible interval from the posterior sampler of \citet{Giannone2015}.

\end{itemize}

\end{enumerate}

\paragraph{Results.}
As in \cref{sec:biasvar_practice} we present simulation results for 200 stationary and 200 non-stationary DGPs, for observed shock identification and averaging across monetary and fiscal policy shocks, with 100 DGPs for each. Coverage results separately by type of shock and for recursive identification are similar, and relegated to \cref{app:fiscal_monetary,app:recursive}. Coverage is approximated by averaging over 1,000 Monte Carlo simulations per DGP.

\begin{figure}[tp!]
\centering
{\textsc{Observed shock: Confidence interval coverage}} \\ 
\vspace{1\baselineskip}
\begin{subfigure}[c]{0.49\textwidth}
\includegraphics[width=\textwidth]{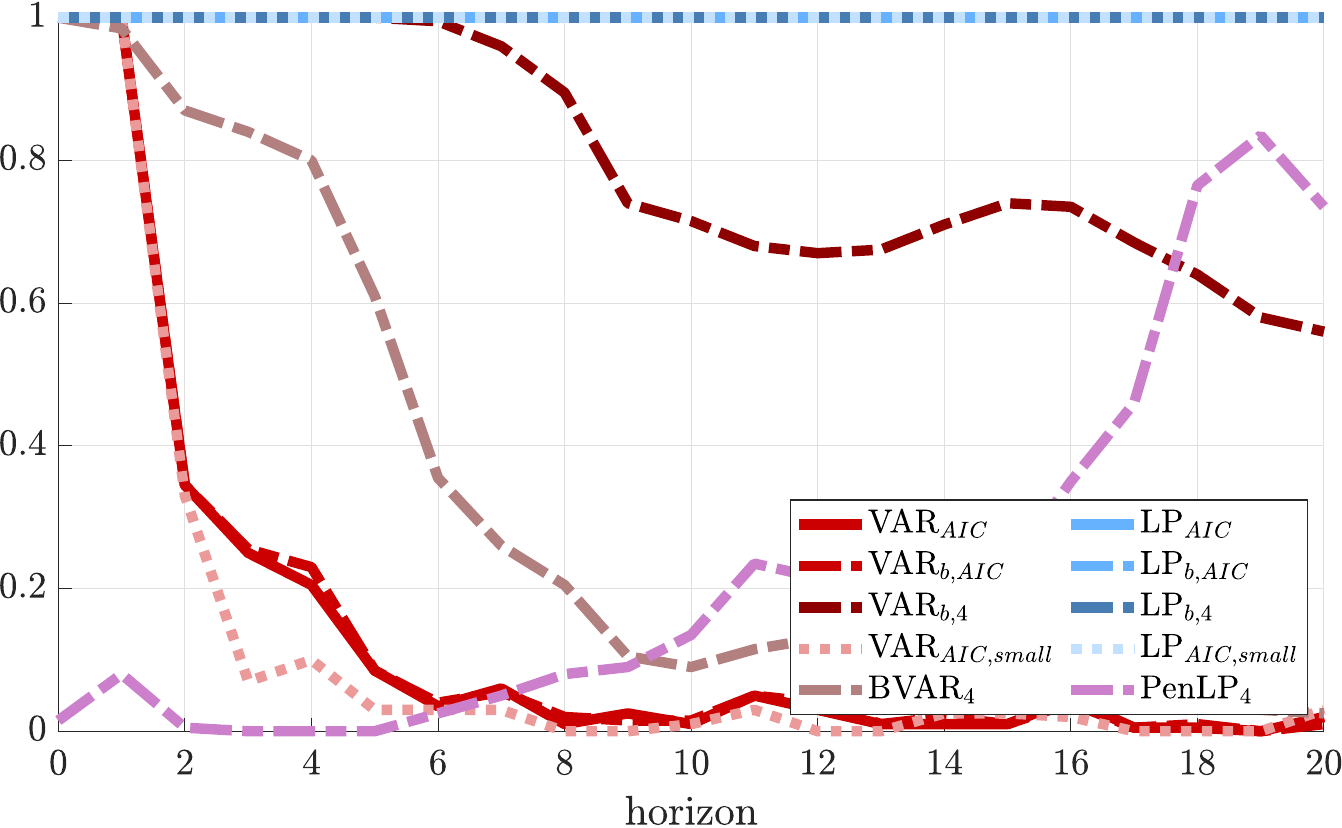}\vspace{0.2cm}
\caption*{\textsc{Stationary DGPs}} \vspace{0.4cm}
\end{subfigure}
\begin{subfigure}[c]{0.49\textwidth}
\includegraphics[width=\textwidth]{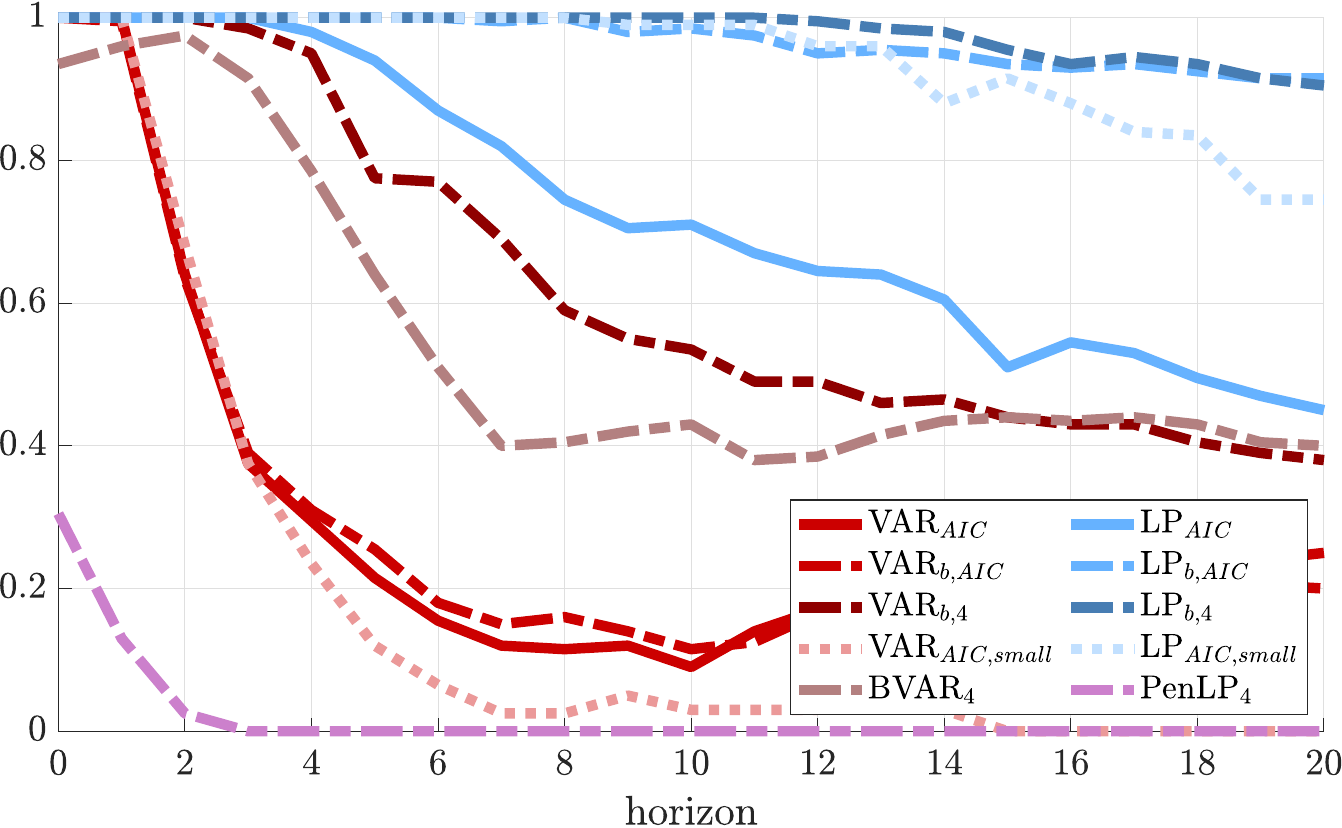}\vspace{0.2cm}
\caption*{\textsc{Non-Stationary DGPs}} \vspace{0.4cm}
\end{subfigure}
\caption{Fraction of DGPs (both stationary and non-stationary) for which the confidence interval coverage probability exceeds 80\%, by inference procedure. The ``$b$'' subscript in the figure legend indicates bootstrap confidence intervals, ``AIC'' indicates lag length selection via the AIC, ``4'' indicates four lags, ``small'' indicates a small system containing only shock and outcome of interest, ``BVAR'' indicates Bayesian VAR, and ``PenLP'' indicates penalized LP.}
\label{fig:covg_obsshock}
\end{figure}

As expected from the theoretical discussions in \cref{subsec:bias_coverage,subsec:longrun_ci}, LPs tend to provide robust uncertainty assessments, while VARs and shrinkage techniques do not. To establish this, \cref{fig:covg_obsshock} shows the fraction of DGPs at which any given inference method delivers confidence intervals with coverage probability above 80\% (and thus close to the target level of 90\%), with VARs in red and LPs in blue.\footnote{\cref{app:simulations_results} furthermore reports average coverage and interval width.} Both analytical and bootstrap LP confidence intervals attain accurate coverage levels for a large fraction of DGPs at all horizons, with the important exception that the bootstrap confidence interval is the only one that works well for non-stationary DGPs at long horizons. By contrast, for the clear majority of DGPs that we consider here, VARs with AIC-selected lag length have a coverage probability below 80\% at all horizons $h>2$. The coverage of longer-lag, bootstrapped VARs is somewhat better, but we see that the coverage of even the best-performing VAR confidence interval is substantially worse than that of the LP bootstrap confidence intervals. The same conclusions extend to shrinkage methods: both Bayesian VARs and penalized LPs undercover severely. That is, even though most of our DGPs feature quite smooth impulse response functions, the shrinkage procedures nevertheless suffer from non-negligible bias, as documented in \cref{subsec:biasvariance_simul}, which has an outsize effect on coverage.

\begin{lesson}

In empirically calibrated simulations, only LPs or (equivalently) VARs with very large lag length deliver robust uncertainty assessments. The best performance is attained with bootstrap confidence intervals.

\end{lesson}

To link back to the discussion in \cref{sec:biasvar_practice}, recall that \citet{Li2024} found in their simulations that a preference for LP over VAR requires an overwhelming concern for bias over variance. We have seen that a desire for robust assessments of statistical uncertainty endogenously creates such a concern.

\section{Summary of recommendations for applied practice}
\label{sec:recommend}

Based on the 12 lessons we have drawn from the available theory and simulation evidence on the econometric properties of LPs and VARs, we make the following summary recommendations for applied researchers who seek to perform inference on the dynamic causal effects of macroeconomic shocks to policies or fundamentals.

\begin{enumerate}[a)]

\item Neither LPs nor VARs solve any identification problems in and of themselves. Regardless of which method you use, you are projecting on something that is supposedly an economic shock. The first step to any empirical analysis is to be transparent about what this shock represents and to convince the reader that this interpretation is sensible.

\item The choice between LPs and VARs amounts to choosing between two different finite-sample estimators of the exact same large-sample estimand. LPs have low bias at the expense of high variance, while VARs (with small to moderate lag length) rely on extrapolation to produce low-variance impulse response estimates at the expense of potentially large bias. The choice of estimation method is likely to matter more the longer is the impulse response horizon of interest.

\item Since plain LP is semiparametrically efficient, \emph{any} impulse response estimator that has lower variance than LP in large samples must be imposing additional restrictions on the transmission mechanism or the identification of the shocks. These restrictions should be explicitly acknowledged.

\item In applications where the goal is to understand what we can learn from the data about dynamic causal effects, uncertainty assessments are a central part of the analysis. There are only two known procedures that provide confidence intervals with accurate coverage across a wide range of empirically relevant DGPs: LPs, as well as VARs with a very large number of lags.

\item If you report VAR confidence intervals, then the estimation lag length should be chosen so large that the confidence intervals approximately coincide with LP intervals. That is, VAR confidence intervals should not be reported without an accompanying LP robustness check. The AIC or other conventional selection criteria will not select sufficiently large lag length to robustify the VAR confidence interval, and conventional model specification tests do not guard sufficiently against the deleterious effects of dynamic misspecification for VAR inference.

\item In applications where the goal is merely to forecast well or to report a point estimate of a causal effect that serves as an input into a policy analysis, frequentist or Bayesian VAR estimators with moderate lag length are attractive since they achieve lower MSE than LP in typical DGPs. However, the associated  confidence intervals cannot be trusted unless the lag length is much larger than has hitherto been common in applied work (and in the case of Bayesian analysis, much less shrinkage should be employed than is conventional for Minnesota-style priors). One possible pragmatic compromise is to report a frequentist or Bayesian VAR point estimate based on a specification with moderate lag length, coupled with a LP confidence interval.\footnote{\cref{app:var_outside} discusses the probability of the event that the VAR estimate falls outside the LP interval.}

\item If using LPs, the choice of controls is just as important as in VAR analysis. One should at least control for (i) variables that are central to the economic identification argument, (ii) several lags of the outcome and impulse variables, and (iii) other strong predictors of either of the latter two variables. See \cref{sec:lpvar_details} for the specific procedure that we recommend for selecting the set of controls and the number of lags for LP.

\item The LP estimator should be bias-corrected using the procedure of \citet{Herbst2024}. Instead of reporting HAC standard errors for LP, simply report the conventional heteroskedasticity-robust standard errors---or even better, the bootstrap confidence interval discussed in \cref{app:lp_var_details}.

\end{enumerate}

\clearpage

\phantomsection
\addcontentsline{toc}{section}{References}
\bibliography{ref}

\end{document}

\title{Online Appendix for ``Local Projections or VARs? A Primer for Macroeconomists''}
\author{\begin{tabular}{ccc}
Jos\'{e} Luis Montiel Olea && Mikkel Plagborg-M{\o}ller \\
Eric Qian && Christian K. Wolf
\end{tabular}}
\date{May 22, 2025}
\maketitle

\bigskip

\noindent This online appendix contains supplemental material on: the comparison of LP and VAR estimation results in prior empirical work (\cref{app:varlp_emp}); implementation details for the LP and VAR estimators that we consider in our simulations (\cref{app:lp_var_details}); a detailed description of our simulation study set-up (\cref{app:simulations_setup}); further simulation results (\cref{app:simulations_results}); theoretical and simulation-based discussion of the probability of the event that the VAR point estimate falls outside of the LP confidence interval (\cref{app:var_outside}).

\begin{appendices}
\crefalias{section}{sappsec}
\crefalias{subsection}{sappsubsec}
\crefalias{subsubsection}{sappsubsubsec}

\section{VARs vs.\ LPs in empirical work}
\label{app:varlp_emp}

We describe how we construct the point estimate and standard error comparison of LPs and VARs in existing applied work in \cref{fig:lp_vs_var_ramey}. Our implementation closely follows our earlier work in \citet[][Online Appendix C]{Montiel2024}, which in turn is based on the literature summary in \citet{Ramey2016}.

We consider four applications in which the researcher has access to a direct measure of a structural shock: to monetary policy, taxes, government spending, and technology. She then estimates the dynamic causal effects of these macroeconomic shocks using either LP or the equivalent (internal-instrument) recursive VAR. The choice of shocks, outcomes, controls, and lags is exactly the same as in our earlier work, as is the computation of standard errors. Overall we obtain LP and VAR impulse response point estimates and standard errors for 385 impulse responses, across all shocks, outcome variables, and horizons. For each we compute standard error ratios and point estimate differences, scaled by the VAR standard error. We finally split the impulse responses into short horizons ($\leq$ one year, 84 observations overall) and long horizons ($>$ one year, 301 observations overall), and report our results for standard error ratios and scaled point estimate differences as boxplots.

\section{Estimation method details}
\label{app:lp_var_details}

We provide implementation details for the LP and VAR estimator variants that we consider in our simulations in \cref{subsec:biasvariance_simul,subsec:inference_simul}.

\subsection{LPs}

Recall from \cref{sec:identification} that the LP estimator of the horizon-$h$ impulse response is the coefficient $\hat{\beta}_h$ in the $h$-step-ahead regression
\begin{equation}
y_{t+h} = \hat{\mu}_h + \hat{\theta}_h^{\text{LP}} x_t + \hat{\gamma}_h' r_t + \sum_{\ell=1}^p \hat{\delta}_{h,\ell}' w_{t-\ell} + \hat{\xi}_{h,t}. \label{eq:lp_details}
\end{equation}
For our different structural identification schemes, the outcome, impulse, and control variables are selected as follows:

\begin{enumerate}

\item {\it Observed shock.} $x_t$ is the observed shock $\varepsilon_{1,t}$, there are no contemporaneous controls $r_t$, and the outcome variable $y_t$ is selected at random from the observed data series. The lagged controls are either the shock plus the full five-dimensional vector of observables, or only the shock plus the outcome of interest.

\item {\it Recursive identification.} $x_t$ is the policy variable (i.e., either the federal funds rate or government purchases), $r_t$ contains all the variables ordered before $x_t$ in the structural shock identification scheme, and $w_t$ always contains the full vector of observables, consistent with the invertibility assumption.

\end{enumerate}

We consider the bias-corrected version of the LP estimator.  We follow \citet{Herbst2024} for the bias correction, using their approximate analytical bias formula for LPs with controls.\footnote{We substitute population autocovariances with sample analogues. We implement an iterative bias correction described in Equation 11 of \citet{Herbst2024}, with the impulse response estimate at horizon $h$ bias-corrected using the previously corrected impulse response estimates at horizons $1,2,\dots,h-1$.} The lag length $p$ is either set exogenously (in our simulations typically to $p = 4$) or selected using the Akaike Information Criterion (AIC) applied to a reduced-form VAR in the observed variables. We use Eicker-Huber-White standard errors, and a modification of the bootstrap routine for lag-augmented local projections suggested in \citet{MontielOlea2021}. An important difference is that instead of using the wild autoregressive bootstrap design, we use the autoregressive residual block bootstrap of \citet{Brueggemann2016}. As implied by the results of \citet{Brueggemann2016}, this is important in order to conduct inference on \emph{structural} impulse responses (rather than reduced-form impulse responses). A description of the algorithm is as follows:

\begin{enumerate}
    \item \label{itm:boot_lp} Compute the LP-based structural impulse response estimate of interest and its standard error.    
    \item \label{itm:boot_var} Estimate the VAR($p$) model by OLS. Compute the corresponding VAR residuals $\hat{u}_t$. Bias-adjust the VAR coefficients using the formula in \citet{Pope1990}.
    \item \label{itm:var_IRF}Compute the structural impulse response of interest implied by the VAR model.
     \item For each bootstrap iteration $b=1,\dots,B$:
    \begin{enumerate}[i)]
        \item Generate properly centered bootstrap residuals $\hat{u}_t^*$ using steps 2 and 3 of the residual-based moving block bootstrap scheme in \cite{Brueggemann2016}, as described on p.\ 73 of their paper (see their Section 4). The block size is chosen using the rule described on p.\ 2665 of \cite{Jentsch2019}.\footnote{The rule gives a block size of 20 for our sample size of $T=240$.}
        \item Draw a block of $p$ initial observations $(y_1^*,\dots,y_p^*)$ uniformly at random from the $T-p+1$ blocks of $p$ observations in the original data.
        \item Generate bootstrap data $y_t^*$, $t=p+1,\dots,T$, by iterating on the bias-corrected VAR($p$) model estimated in step \ref{itm:boot_var}, using the innovations $\hat{u}_t^*$.
        \item Apply the LP estimator to the bootstrap data $\lbrace y_t^* \rbrace$ and compute its Eicker-Huber-White standard error. 
         
         \item Store the t-statistic for the LP estimate, making sure that the statistic is centered around the VAR-implied structural impulse response from step \ref{itm:var_IRF}. As explained by \citet{MontielOlea2021}, it is critical that the bootstrap t-statistic is centered at the VAR-implied impulse response (see step 4(v) on p.\ 1808 of that paper), not the LP estimate from step \ref{itm:boot_lp}.
    \end{enumerate}
    \item Compute the $a/2$ and $1-a/2$ quantiles of the $B$ draws of the t-statistic, $b=1,\dots,B$. %Denote these by $\hat{Q}_{\alpha/2}$ and $\hat{Q}_{1-\alpha/2}$, respectively.
    \item Return the percentile-t bootstrap confidence interval \citep[e.g.,][step 6, p.\ 1808]{MontielOlea2021}.
\end{enumerate}

We also consider a shrinkage variant of LP, the penalized LP of \citet{Barnichon2019}. As suggested by those authors, we model impulse responses using B-spline basis functions, and penalize deviations from a quadratic function of the impulse response horizon $h$, up to $20$, with the penalty parameter selected using 5-fold cross-validation. For confidence intervals we similarly follow the suggestions of the authors, using their heuristic procedure (see p.\ 525), with the exact same undersmoothing to reduce the bias induced by shrinkage.

\subsection{VARs}

Recall from \cref{sec:identification} that the VAR impulse response estimator is based on the reduced-form VAR
\begin{equation}
w_t = \hat{c} + \sum_{\ell=1}^p \hat{A}_\ell w_{t-\ell} + \hat{u}_t. \label{eq:var_details}
\end{equation}
with $\var(u_t) = \hat{B} \hat{B}'$ where $\hat{B}$ is the Cholesky decomposition of the estimated forecast error variance-covariance matrix. The two identification schemes are implemented as follows:

\begin{enumerate}

\item {\bf Observed shock.} $w_t$ contains the observed shock as well as either all other five observed series, or only the outcome variable of interest. The observed shock is ordered first in the recursive orthogonalization of the reduced-form innovations.

\item {\bf Recursive identification.} $w_t$ consists of the five observed series, ordered as indicated in our discussion of the structural monetary and fiscal shock identification schemes (see \cref{app:dfm_dgps_estimands}). We do not consider a small version of this system, consistent with the invertibility assumption.

\end{enumerate}

The reduced-form VAR coefficient matrices are estimated using the analytical bias correction of \citet{Pope1990}, following the recommendations in \citet{Kilian1998}. The lag length $p$ is either set exogenously (in our simulations typically to $p = 4$) or selected using the AIC for \eqref{eq:var_details}. We use Eicker-Huber-White standard errors, and bootstrap VAR estimates using the residual block bootstrap of \cite{Brueggemann2016}. The block size is chosen using the rule of thumb of \citet[][p.\ 2665]{Jentsch2019}, giving a block size of 20 for our sample size of $T=240$.  Following the recommendation of \citet{Inoue2020}, we report the Efron bootstrap confidence interval.

For shrinkage, we estimate a Bayesian VAR using the default prior recommendations of \citet{Giannone2015}, largely following their replication code, though with some minor adjustments as discussed in \citet[][Appendix B]{Li2024}. We draw 500 times from the posterior, reporting posterior means of the impulse responses (for our bias-variance trade-off plots) and constructing posterior credible intervals using 5th and 95th percentiles (for uncertainty assessments).

\section{Simulation study details}
\label{app:simulations_setup}

\subsection{DFM estimation}

We estimate the encompassing stationary and non-stationary DFMs on the data set of \citet{Stock2016}, proceeding as in \citet{Li2024}, but additionally allowing for ARCH disturbances. For the stationary DFM, we follow the exact same steps as in Online Appendix F.2 (p.\ 16) of \citeauthor{Li2024}, which in turn replicates the original analysis by \citeauthor{Stock2016} as well as in \citet{Lazarus2018}. For the non-stationary DFM, we first transform variables, then select the number of factors and lags, and finally estimate the factor VECM just as in Online Appendix C (pp.\ 2--4) of \citeauthor{Li2024}.

To better capture likely challenges for inference in applied practice, we generalize the estimated DFMs by allowing for heteroskedastic errors. Specifically, given the DFM estimated in the first step, we next estimate separate ARCH(1) models for the \emph{reduced-form} residuals in the factor and idiosyncratic equations. For the simulation DGP, we conservatively model the \emph{structural} shocks to the factors as following independent ARCH(1) models with ARCH coefficient equal to the maximum of the estimated reduced-form factor ARCH coefficients; for the idiosyncratic innovations we just directly use the estimated actual ARCH coefficients, censored above at 0.7 (this censoring affects only a handful of series and is consistent with the estimated confidence intervals).

\subsection{DGP selection and impulse response estimands}
\label{app:dfm_dgps_estimands}

We draw our individual DGPs from the two encompassing DFMs by proceeding as follows. For all DGPs, we restrict attention to the following 17 oft-used series (with \citeauthor{Stock2016} Data Appendix series \# in brackets): \emph{real GDP (1); real consumption (2); real investment (6); real government expenditure (12); the unemployment rate (56); personal consumption expenditure prices (95); the GDP deflator (97); the core consumer price index (121); average hourly earnings (132); the federal funds rate (142); the 10-year Treasury rate (147); the BAA 10-year spread (151); an index of the U.S. dollar exchange rate relative to other major currencies (172); the S\&P 500 (181); a real house price index (193); consumer expectations (196); and real oil prices (202).} We then draw several random combinations of five series from this overall set of salient time series, subject to the constraint that each DGP contains either the federal funds rate or government spending (for monetary or fiscal shock estimands, respectively, as discussed further below) as well as at least one real activity series (categories 1–3 in the \citeauthor{Stock2016} data appendix) and one price series (category 6). For both the stationary and non-stationary DFM we draw 100 monetary and 100 fiscal DGPs in this way.

\paragraph{Observed shock identification.}
We define the monetary policy shock as the (unique) linear combination of the innovations in the factor equation that maximizes the impact impulse response of the federal funds rate, and analogously for the fiscal policy shock, with government spending as the maximized response variable. We then assume that the econometrician directly observes this shock of interest, together with the five observables drawn from the list of salient time series, as we discussed above. She estimates the propagation of the shock using either LPs with the shock as the impulse variable (and no contemporaneous controls) or a recursive VAR with the observed shock ordered first. The outcome of interest is randomly selected among the observable series, not including the fiscal or monetary policy instruments (i.e., government purchases or the federal funds rate).

We also estimate LP and VAR specifications that use a smaller set of observables. Here the system only contains the observed shock as well as the outcome variable of interest.

\paragraph{Recursive identification.}
For recursive identification, the researcher only observes the five time series drawn from the encompassing DFM. We then define as the object of interest impulse responses with respect to a recursive orthogonalization of the reduced-form (Wold) forecast errors in the VAR($\infty$) representation of the observables. For monetary policy, we order the federal funds rate last and then call the orthogonalized innovation to that variable a monetary policy shock, restricting all other variables to not respond contemporaneously to monetary policy, as in \citet{Christiano1999}. For fiscal policy, we order government spending first and then call the innovation to that variable a fiscal policy shock, thus restricting the fiscal authority to respond to all of the other innovations with a lag, following \citet{Blanchard2002}. %The overall identifying assumptions here are thus invertibility together with particular temporal orderings.
See \citet[][Online Appendix D]{Li2024} for further details.

\subsection{Population estimands}
\label{app:simulations_estimands}

For our visual illustration of LP-VAR equivalence in \cref{fig:lp_vs_var_dfm} we consider one particular monetary policy DGP from the stationary DFM, with recursive shock identification. The observables in our system are (\citeauthor{Stock2016} Data Appendix series \# in brackets): the unemployment rate (56); real GDP (1); the core consumer price index (121); the BAA 10-year spread (151); and the federal funds rate (142). The outcome of interest is unemployment. We then simulate a large sample, and use recursive LPs and VARs with $p \in \{ 2, 6, 12 \}$ lags to estimate the propagation of the recursively identified monetary shock, as discussed above. We compare these finite-lag LP and VAR estimands with the true population projection on the recursively identified monetary policy innovation, which we estimate using a numerical approximation to an infinite-lag VAR.

\subsection{Degree of misspecification}
\label{app:simulations_misspec}

Given a DGP---i.e., a five-variable (for recursive identification) or six-variable (for observed shock identification) system randomly drawn from the encompassing DFM---and a lag length $p$, we can represent that DGP as a VARMA($p$,$\infty$), following the same steps as those outlined in \citet[][Footnote 8]{Montiel2024}. %For any given sample size $T$ and degree of local mis-specification $\zeta$, we can then map that VARMA($p$,$\infty$) into the locally mis-specified VAR($p$) framework of \citeauthor{Montiel2024}, obtaining in particular the misspecification lag polynomial $\alpha(L)$.
We then obtain the total degree of misspecification as
\begin{equation*}
\sqrt{T \times \mathcal{M}} = \|\alpha(L)\| \equiv \sqrt{\sum_{\ell=1}^\infty \tr\lbrace D \alpha_\ell'D^{-1}\alpha_\ell\rbrace},
\end{equation*}
where $D$ is the variance-covariance matrix of the contemporaneous innovations. Please note that this computation is relevant only for \cref{fn:avg_misspecification}, where we report the average degree of misspecification across our stationary DGPs.

\section{Supplementary simulation results}
\label{app:simulations_results}

We provide several supplementary simulation results to complement our main findings reported in \cref{subsec:biasvariance_simul,subsec:inference_simul}. We here report detailed results for: recursive identification; monetary and fiscal policy shocks considered separately; as well as average confidence interval coverage and width.

\subsection{Recursive identification}
\label{app:recursive}

\begin{figure}[t]
\centering
{\textsc{Recursive identification: bias-variance trade-off}} \\ 
\vspace{0.8\baselineskip}
\begin{subfigure}[c]{0.49\textwidth}
\includegraphics[width=\textwidth]{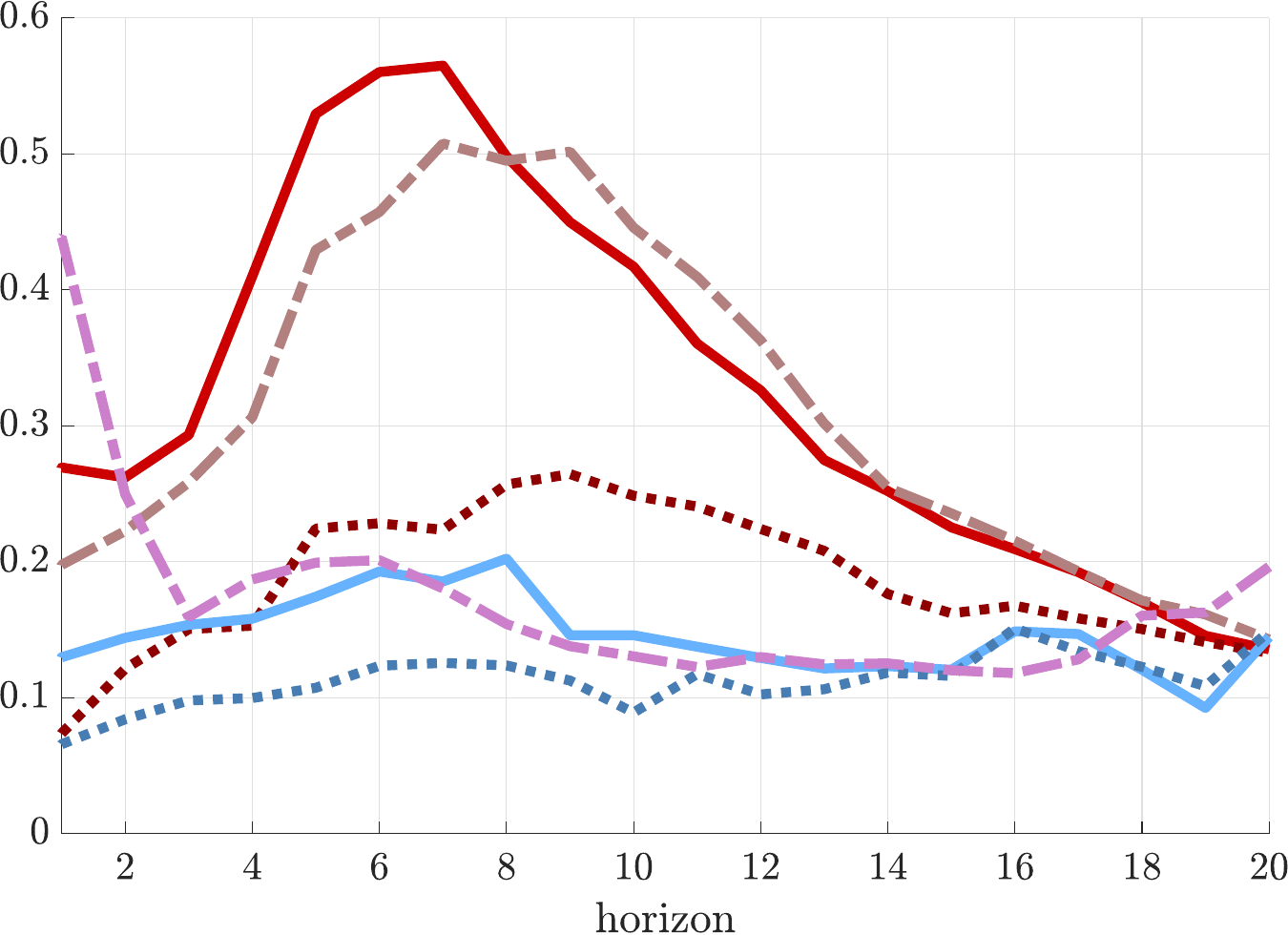}\vspace{0.2cm}
\caption*{\textsc{Bias}} \vspace{0.2cm}
\end{subfigure}
\begin{subfigure}[c]{0.49\textwidth}
\includegraphics[width=\textwidth]{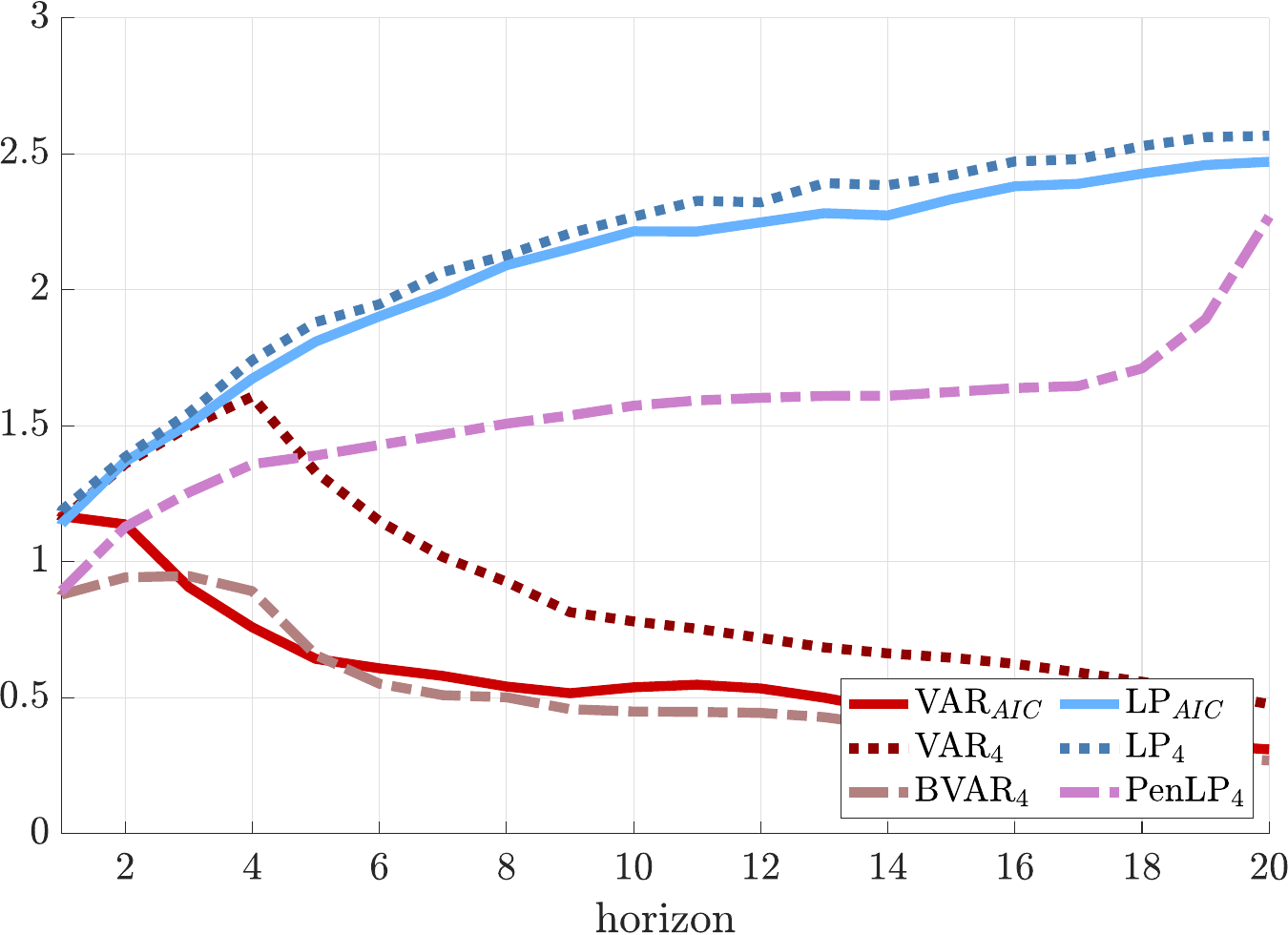}\vspace{0.2cm}
\caption*{\textsc{Standard deviation}} \vspace{0.2cm}
\end{subfigure}
\caption{Median (across DGPs) of absolute bias $\left |\mathbb{E} \left [ \hat{\theta}_h-\theta_h \right ] \right |$ (left panel) and standard deviation $\sqrt{\var(\hat{\theta}_h)}$ (right panel) of the different estimation procedures, relative to $\sqrt{\frac{1}{21}\sum_{h=0}^{20}\theta_h^2}$.}
\label{fig:biasstd_recursive}
\end{figure}

\begin{figure}[t]
\centering
\textsc{Recursive identification: CI coverage} \\ 
\vspace{1\baselineskip}
\includegraphics[width=0.7\linewidth]{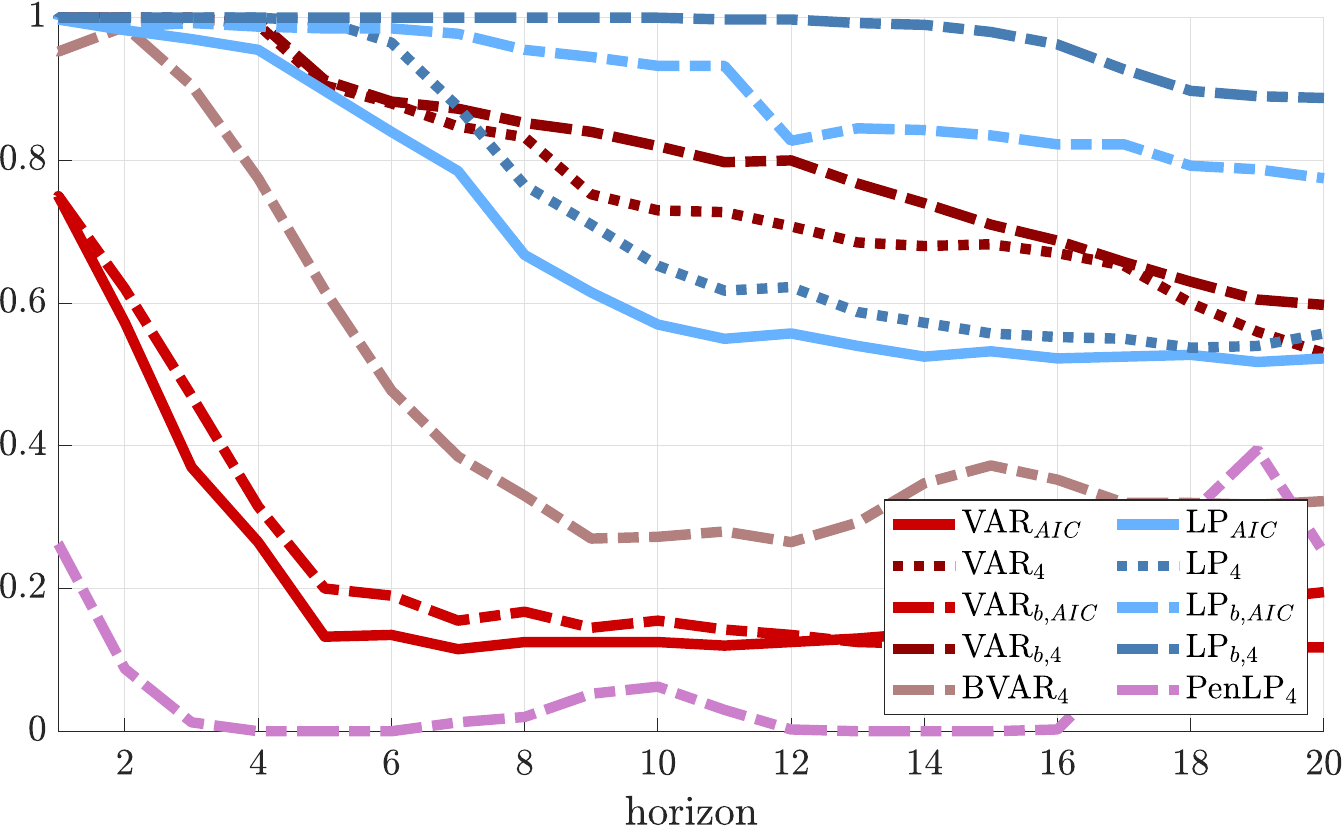}
\caption{Fraction of DGPs (both stationary and non-stationary) for which the confidence interval coverage probability exceeds 80\%. See caption for \cref{fig:covg_obsshock} for an explanation of the abbreviations in the legend.}
\label{fig:covg_recursive}
\end{figure}

While our headline results in \cref{subsec:biasvariance_simul,subsec:inference_simul} are reported for observed shock identification, broadly similar lessons for both the bias-variance trade-off and for inference emerge under recursive shock identification. Visual summaries are provided in \cref{fig:biasstd_recursive,fig:covg_recursive}.

\cref{fig:biasstd_recursive} shows the bias-variance trade-off, now in the interest of space averaging across both monetary and fiscal shocks and across stationary and non-stationary DGPs. We see the same patterns as for observed shock identification: the bias-variance trade-off is stark, long lag lengths for the VAR align it with LPs, and the variance cost of LPs is large. Differently from the observed shock case, we do not report any results for ``small'' specifications, simply because the control vector is now integral to the economic identifying assumptions.

\cref{fig:covg_recursive} displays the coverage of confidence intervals, again across all DGPs. As in the observed shock case, VARs with lag length selected by standard information criteria significantly undercover, while LPs tend to cover well, in particular with bootstrapped confidence intervals. Also as for observed shocks, bootstrapping is particularly important for accurate long-horizon coverage in the non-stationary DGPs. Finally we now also see that inclusion of longer lag lengths is more important for LP, simply because those lags now matter for the structural identification scheme, ensuring that the shock of interest is indeed spanned by the regression residuals.

\subsection{Fiscal and monetary shocks}
\label{app:fiscal_monetary}

\begin{figure}[p]
\centering
{\textsc{Observed fiscal shock: bias-variance trade-off}} \\ 
\vspace{0.8\baselineskip}
\begin{subfigure}[c]{0.49\textwidth}
\includegraphics[width=\textwidth]{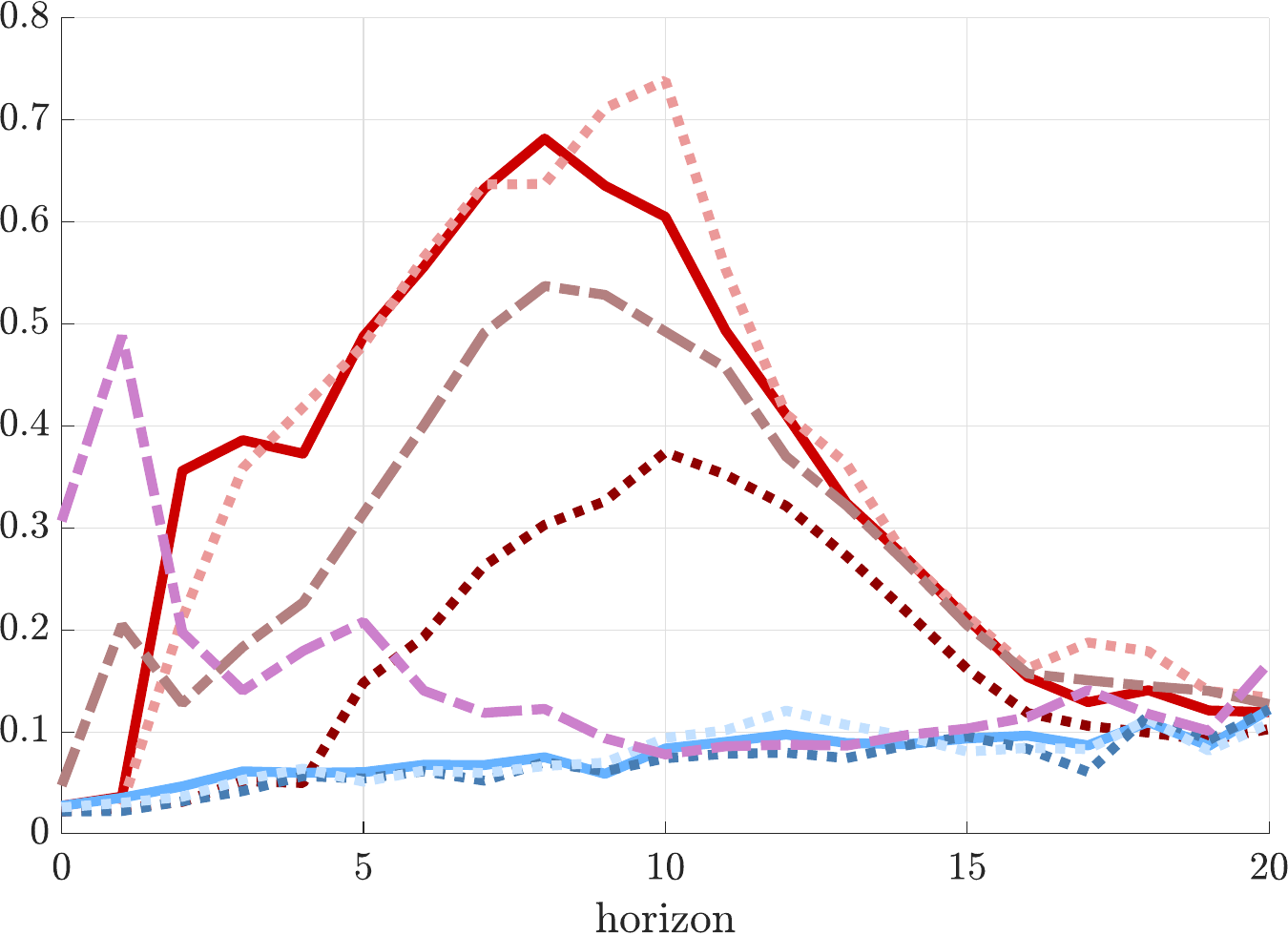}\vspace{0.2cm}
\caption*{\textsc{Bias}} \vspace{0.2cm}
\end{subfigure}
\begin{subfigure}[c]{0.49\textwidth}
\includegraphics[width=\textwidth]{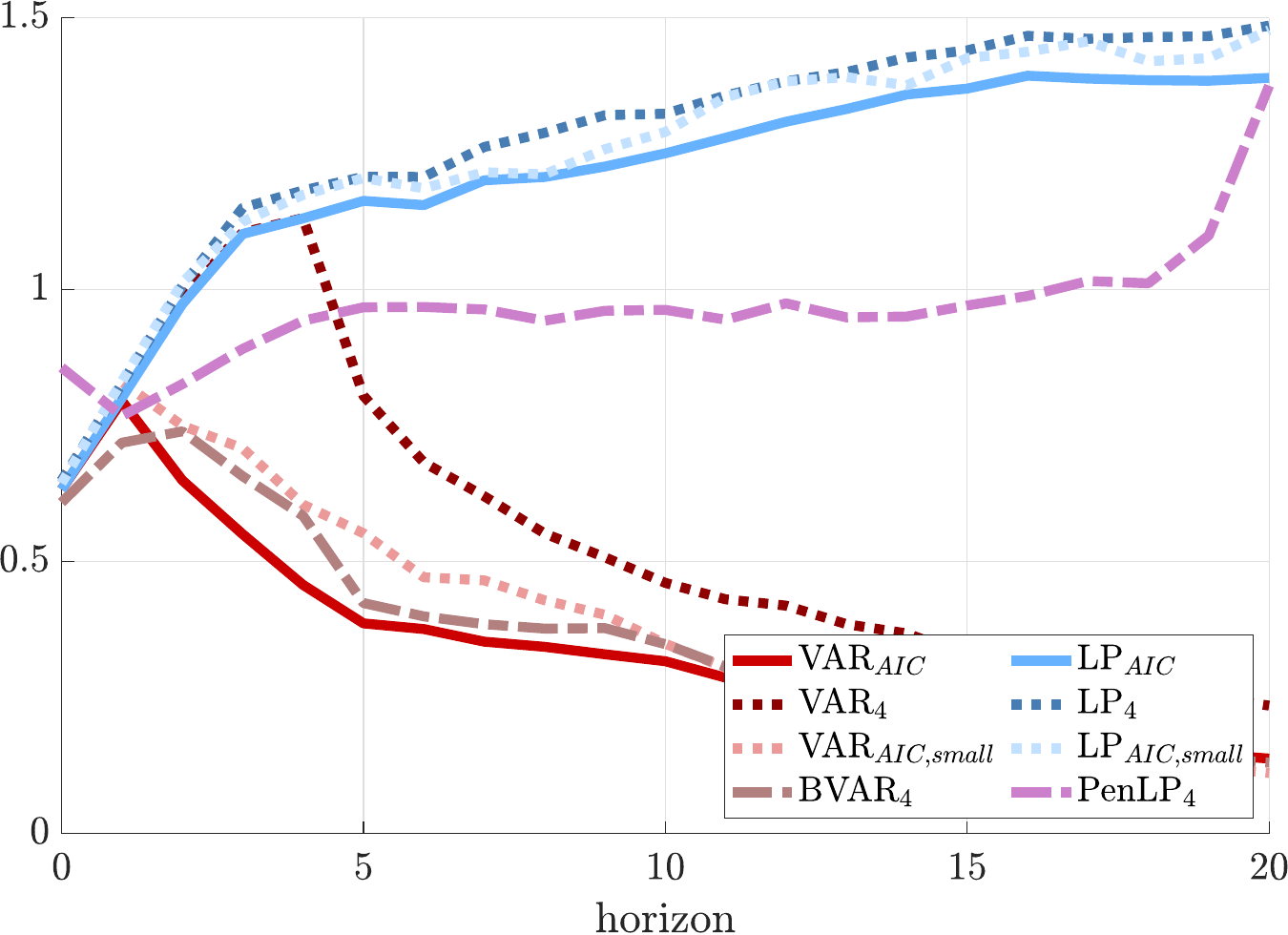}\vspace{0.2cm}
\caption*{\textsc{Standard deviation}} \vspace{0.2cm}
\end{subfigure}
\caption{Median (across DGPs) of absolute bias $\left |\mathbb{E} \left [ \hat{\theta}_h-\theta_h \right ] \right |$ (left panel) and standard deviation $\sqrt{\var(\hat{\theta}_h)}$ (right panel) of the different estimation procedures, relative to $\sqrt{\frac{1}{21}\sum_{h=0}^{20}\theta_h^2}$.}
\label{fig:biasstd_obsshock_g}
\end{figure}

\begin{figure}[p]
\centering
\textsc{Observed fiscal shock: CI coverage} \\ 
\vspace{1\baselineskip}
\includegraphics[width=0.7\linewidth]{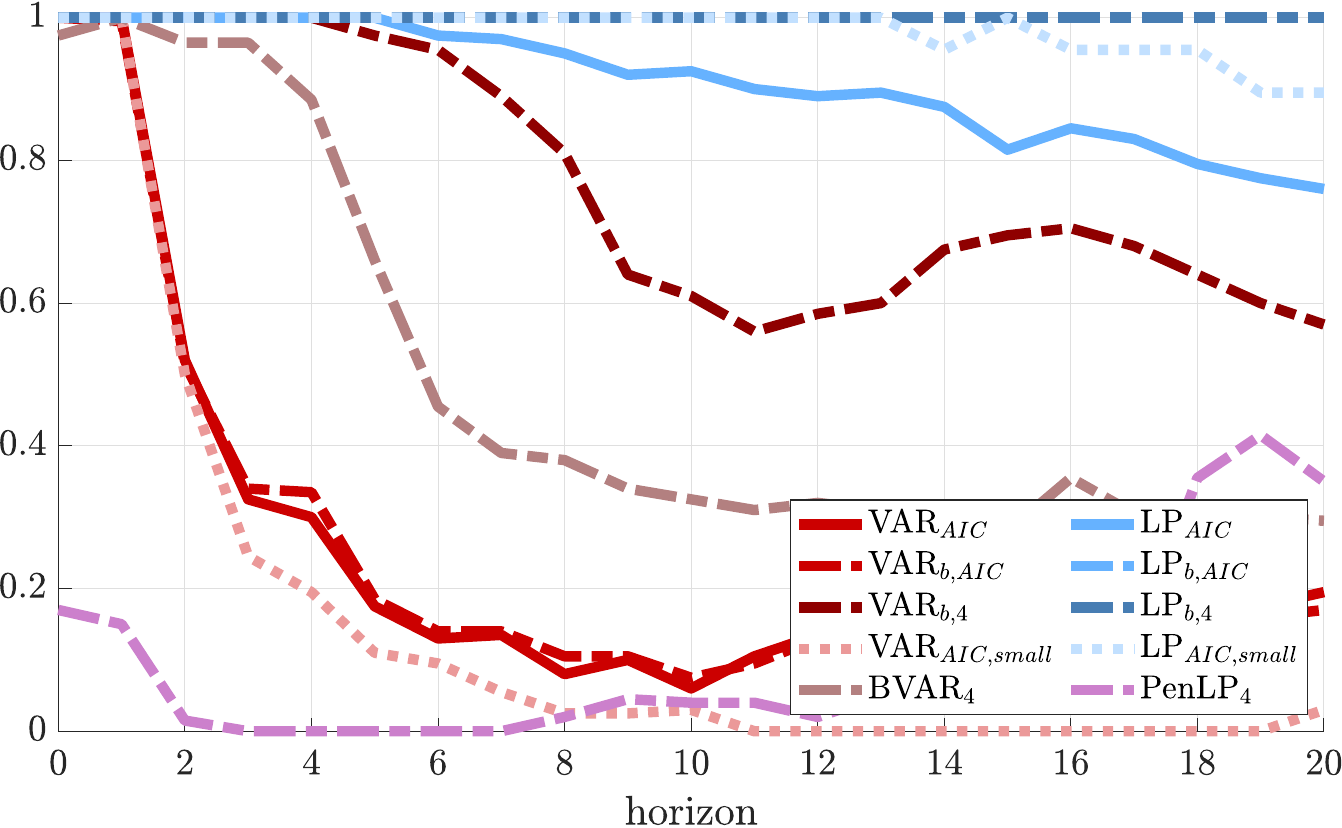}
\caption{Fraction of DGPs (both stationary and non-stationary) for which the confidence interval coverage probability exceeds 80\%. See caption for \cref{fig:covg_obsshock} for an explanation of the abbreviations in the legend.}
\label{fig:covg_obsshock_g}
\end{figure}

Our reported conclusions are not sensitive to the type of (policy) shock that we consider. To establish this, \cref{fig:biasstd_obsshock_g,fig:covg_obsshock_g} show bias, variance, and coverage results for fiscal shocks (combining stationary and non-stationary DGPs), while \cref{fig:biasstd_obsshock_g,fig:covg_obsshock_g} do the same for monetary shocks. The figures by shock echo the messages of our main figures, which average across shocks: there is a meaningful bias-variance trade-off; the variance cost of LPs is high; and only LP methods robustly attain high coverage.

\begin{figure}[p]
\centering
{\textsc{Observed monetary shock: bias-variance trade-off}} \\ 
\vspace{0.8\baselineskip}
\begin{subfigure}[c]{0.49\textwidth}
\includegraphics[width=\textwidth]{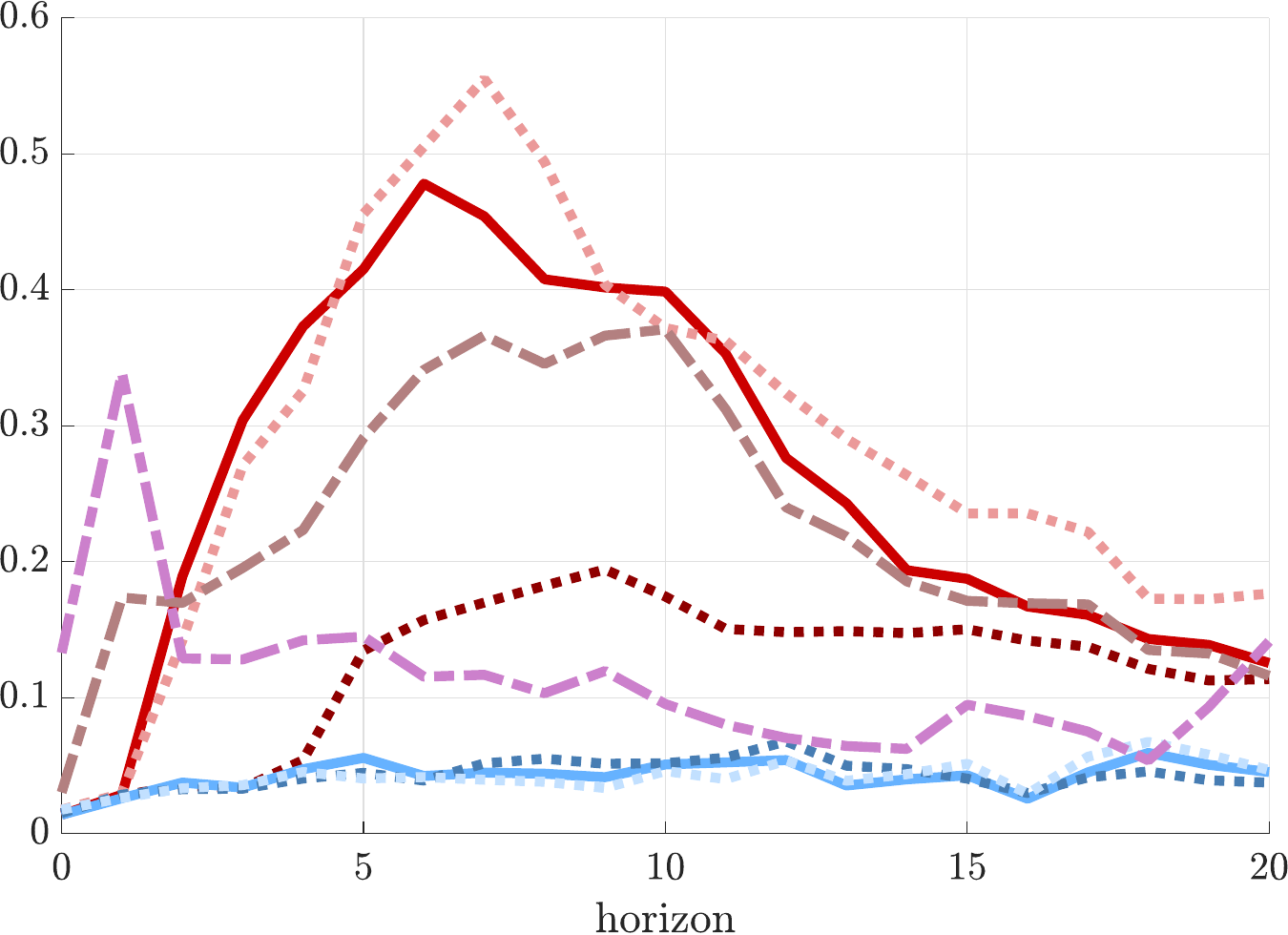}\vspace{0.2cm}
\caption*{\textsc{Bias}} \vspace{0.2cm}
\end{subfigure}
\begin{subfigure}[c]{0.49\textwidth}
\includegraphics[width=\textwidth]{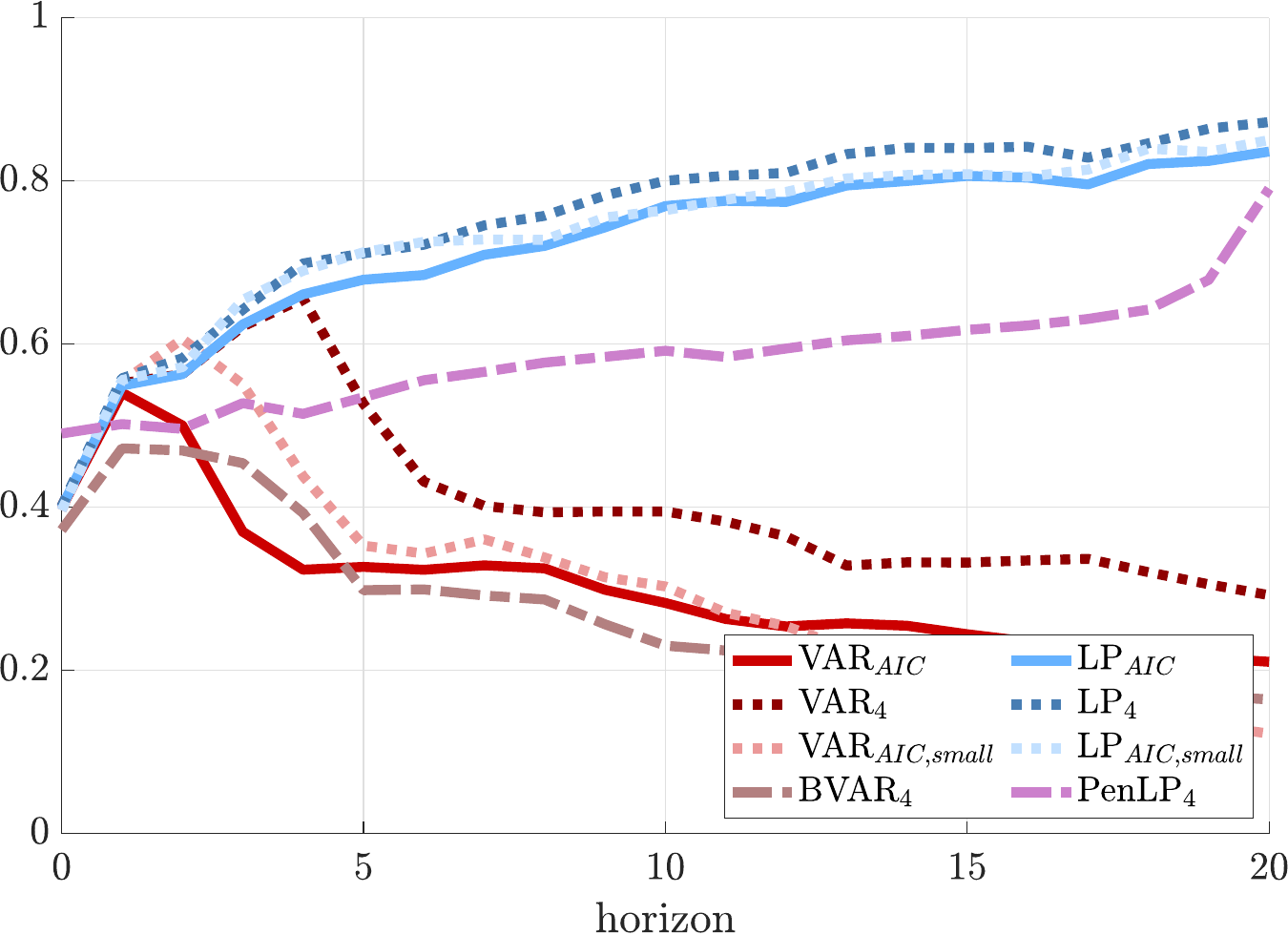}\vspace{0.2cm}
\caption*{\textsc{Standard deviation}} \vspace{0.2cm}
\end{subfigure}
\caption{Median (across DGPs) of absolute bias $\left |\mathbb{E} \left [ \hat{\theta}_h-\theta_h \right ] \right |$ (left panel) and standard deviation $\sqrt{\var(\hat{\theta}_h)}$ (right panel) of the different estimation procedures, relative to $\sqrt{\frac{1}{21}\sum_{h=0}^{20}\theta_h^2}$.}
\label{fig:biasstd_obsshock_mp}
\end{figure}

\begin{figure}[p]
\centering
\textsc{Observed monetary shock: CI coverage} \\ 
\vspace{1\baselineskip}
\includegraphics[width=0.7\linewidth]{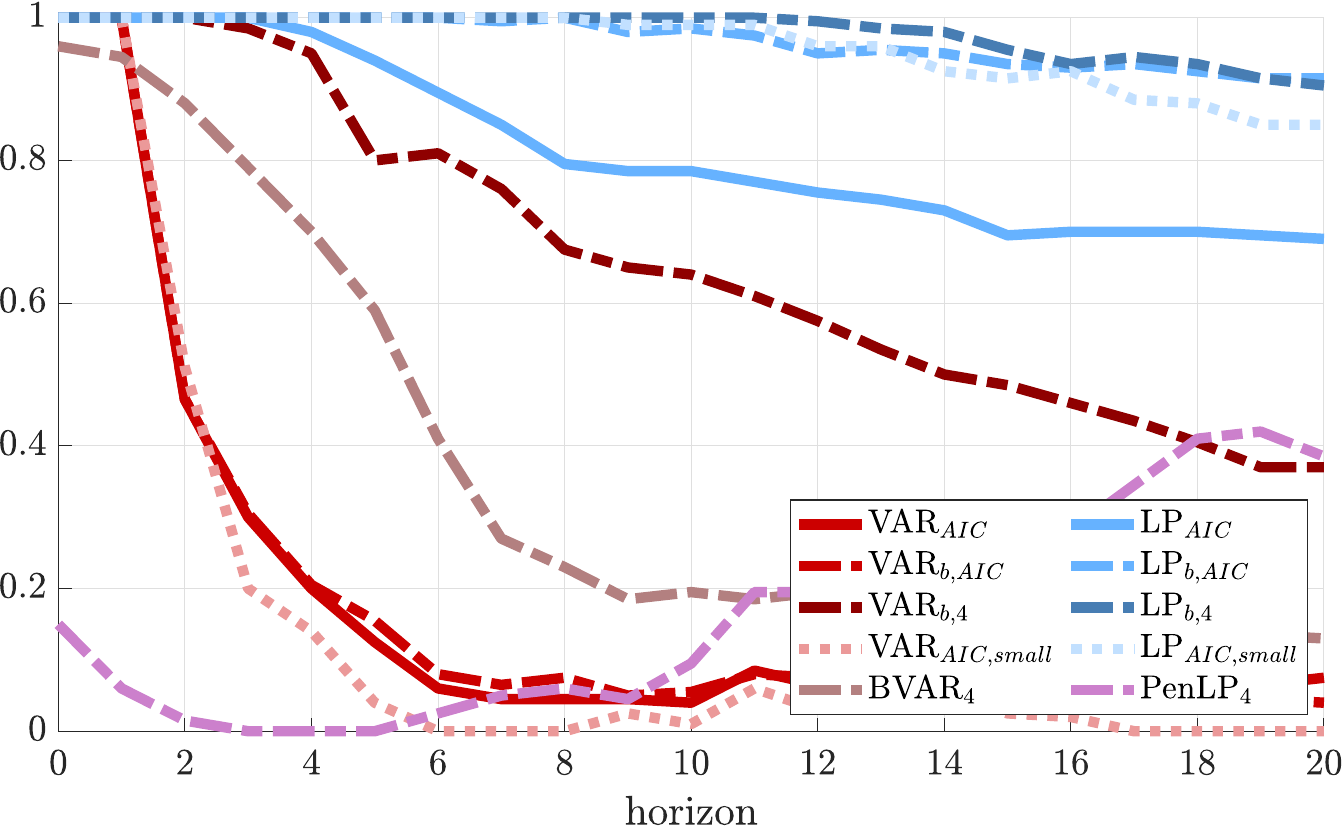}
\caption{Fraction of DGPs (both stationary and non-stationary) for which the confidence interval coverage probability exceeds 80\%. See caption for \cref{fig:covg_obsshock} for an explanation of the abbreviations in the legend.}
\label{fig:covg_obsshock_mp}
\end{figure}

\subsection{Coverage probability and CI width}

\begin{figure}[tp]
\centering
\textsc{Observed shock: CI coverage probability \& width} \\ 
\vspace{1\baselineskip}
\includegraphics[width=\linewidth]{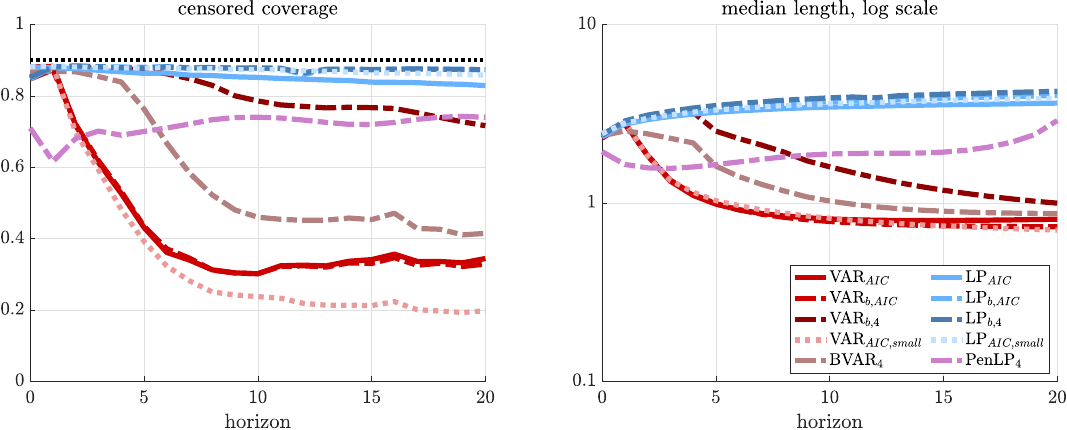}
\caption{Coverage probabilities (left panel, coverage is censored above at 90\%) and median confidence interval length (right panel) for VAR (red) and LP (blue) confidence intervals, averaged across all DGPs (both stationary and non-stationary) separately at each horizon. The right panel normalizes the interval length by the overall magnitude $\sqrt{\frac{1}{21}\sum_{h=0}^{20}\theta_h^2}$ of the true impulse response function prior to averaging across DGPs. See caption for \cref{fig:covg_obsshock} for an explanation of the abbreviations in the legend.}
\label{fig:covgwidth_obsshock}
\end{figure}

While in \cref{subsec:inference_simul} we report the fraction of LP and VAR confidence intervals with coverage above 80\%, \cref{fig:covgwidth_obsshock} here instead shows the coverage probability (left panel) and median confidence interval length (right panel) of our different inference procedures \emph{averaged} across all DGPs, both stationary and non-stationary. Before averaging the coverage probabilities across DGPs, we censor them above at 90\% so that over-coverage is not rewarded. The picture that emerges is yet again consistent with our theoretical and practical messages: VAR and shrinkage confidence intervals can be quite a bit shorter, but this tends to come at the cost of (sometimes material) under-coverage, in particular at longer horizons. The VAR specifications with longer lag lengths yield correct coverage only at those horizons where the VAR confidence intervals are essentially as wide as those of LP.

\section{Probability of the VAR estimate falling outside the LP interval}
\label{app:var_outside}

In the theoretical framework of \cref{subsec:bias_variance_analytics}, the probability that the VAR estimate falls outside the LP confidence interval equals
\begin{align*}
P\left(\left|\hat{\theta}_{h}^{\text{VAR}}-\hat{\theta}_{h}^{\text{LP}}\right|>\tau_{h,\text{LP}}z_{1-a/2}\right) &\approx P\left(\left|N\left(b_h(p),\tau_{h,\text{LP}}^2-\tau_{h,\text{VAR}}(p)^2\right)\right| >\tau_{h,\text{LP}}z_{1-a/2}\right) \\
&= P\left(\left|N\left(\frac{b_h(p)}{\tau_{h,\text{VAR}}(p)},\frac{\tau_{h,\text{LP}}^2}{\tau_{h,\text{VAR}}^2(p)}-1\right)\right| >\frac{\tau_{h,\text{LP}}}{\tau_{h,\text{VAR}}(p)}z_{1-a/2}\right),
\end{align*}
where the approximation becomes exact asymptotically, due to \eqref{eq:large_sample_distr} and the fact that the asymptotic covariance of the VAR and LP estimators equals the variance of the VAR estimator, as shown by \citet{Montiel2024}. Since the right-hand side above is increasing in the bias/standard-deviation ratio of the VAR estimator, the bound \eqref{eq:bias_bound} implies that the probability is no greater than
\begin{align*}
& P\left(\left|N\left(\sqrt{T \times \mathcal{M}} \times \sqrt{\frac{\tau_{h,\text{LP}}^2}{\tau_{h,\text{VAR}}^2(p)}-1},\frac{\tau_{h,\text{LP}}^2}{\tau_{h,\text{VAR}}^2(p)}-1\right)\right| >\frac{\tau_{h,\text{LP}}}{\tau_{h,\text{VAR}}(p)}z_{1-a/2}\right) \\
&= P\left(\left|N\left(\sqrt{T \times \mathcal{M}},1\right)\right| > \frac{z_{1-a/2}}{\sqrt{1-\tau_{h,\text{VAR}}^2(p)/\tau_{h,\text{LP}}^2}} \right).
\end{align*}
When $1-a=90\%$ and $\tau_{h,\text{VAR}}(p)/\tau_{h,\text{LP}}=0.4$, then the above probability bound equals 21.6\% when $\sqrt{T \times \mathcal{M}}=1$, and 58.1\% when $\sqrt{T \times \mathcal{M}}=2$. Hence, there exist DGPs with a moderate amount of misspecification for which it is quite likely that the LP interval does not contain the VAR estimate.

\begin{figure}[tp]
	\centering
	\textsc{Observed shock: VAR point estimates in LP confidence intervals} \\
	\vspace{1\baselineskip}
	\includegraphics[width=0.6\linewidth]{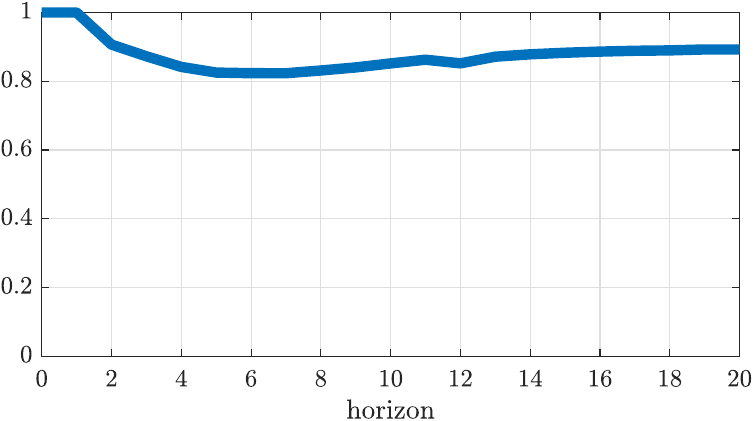}
	\caption{Probability that the VAR point estimates are contained in the LP percentile-t bootstrap confidence intervals, averaged across all DGPs (stationary and non-stationary, monetary and fiscal), by impulse response horizon. The VAR and LP lag lengths are selected using the AIC.}
	\label{fig:varinlpci_obsshock}
\end{figure}

However, in our empirically calibrated simulation study in \cref{subsec:inference_simul}, the clear majority of VAR point estimates do lie inside the LP confidence interval. \cref{fig:varinlpci_obsshock} illustrates, plotting the share of VAR point estimates that are inside the LP interval by impulse response horizon $h$, averaged over all (stationary and non-stationary, monetary and fiscal) DGPs. The share is throughout in excess of 80 per cent, and particularly high at short horizons, consistent with our theoretical discussion.

\clearpage
\phantomsection
\addcontentsline{toc}{section}{References}
\bibliography{ref}

\end{appendices}